\documentclass[12pt]{article}
\usepackage{latexsym}
\usepackage{float}
\usepackage{amsmath,bm,amssymb}
\usepackage{colordvi}
\usepackage{multicol,multirow}
\usepackage{color}
\usepackage{rotating}
\usepackage{url,hyperref}
\usepackage[ruled]{algorithm2e}
\usepackage{pdflscape}
\usepackage{subfigure}
\usepackage{tabu}
\usepackage{multirow}
\usepackage{booktabs}
\usepackage{graphicx}
\usepackage[utf8]{inputenc}
\usepackage{chngcntr}

\usepackage[super,sort&compress]{natbib}
\setcitestyle{super}

\def\doublespace{\baselineskip=22pt}
\usepackage{lscape}

\begin{document}
	\doublespace
	\baselineskip 2.8ex

\begin{center}
	{\bf\LARGE Semi-parametric Bayesian variable selection for gene-environment interactions}
	
	\vspace{1.6em}
	
	{\bf Jie Ren$^{1}$, Fei Zhou$^{1}$,  Xiaoxi Li$^{1}$, Qi Chen$^2$, Hongmei Zhang$^3$, Shuangge Ma$^4$, Yu Jiang$^3$ and Cen Wu$^{\ast1}$}
	
	{ $^1$ Department of Statistics, Kansas State University, Manhattan, KS}\\
	
	{ $^2$ Department of Pharmacology, Toxicology and Therapeutics, University of Kansas Medical Center, Kansas City, KS}\\
	
	{ $^3$ Division of Epidemiology, Biostatistics and Environmental Health, School of Public Health, University of Memphis, Memphis, TN} \\
	
	{ $^4$ Department of Biostatistics, Yale University, New Haven, CT}\\
	
\end{center}

\noindent{\bf *Corresponding author}: Cen Wu, wucen@ksu.edu

\section*{Abstract}
Many complex diseases are known to be affected by the interactions between genetic variants and environmental exposures beyond the main genetic and environmental effects. Study of gene-environment (G$\times$E) interactions is important for elucidating the disease etiology. Existing Bayesian methods for G$\times$E interaction studies are challenged by the high-dimensional nature of the study and the complexity of environmental influences. Many studies have shown the advantages of penalization methods in detecting G$\times$E interactions in ``large p, small n'' settings. However, Bayesian variable selection, which can provide fresh insight into G$\times$E study, has not been widely examined. We propose a novel and powerful semi-parametric Bayesian variable selection model that can investigate linear and nonlinear G$\times$E interactions simultaneously. Furthermore, the proposed method can conduct structural identification by distinguishing nonlinear interactions from main-effects-only case within the Bayesian framework. Spike and slab priors are incorporated on both individual and group levels to identify the sparse main and interaction effects. The proposed method conducts Bayesian variable selection more efficiently than existing methods. Simulation shows that the proposed model outperforms competing alternatives in terms of both identification and prediction. The proposed Bayesian method leads to the identification of main and interaction effects with important implications in a high-throughput profiling study with high-dimensional SNP data.

\noindent{\bf Keywords:}Bayesian variable selection, Gene-environment interactions, High-dimensional genomic data, Semi-parametric modeling, MCMC

\section{Introduction}
\label{makereference3.1}

It has been widely recognized that the genetic and environmental main effects alone are not sufficient to decipher an overall picture of the genetic basis of complex diseases. The Gene-Environment (G$\times$E) interactions also play vital roles in dissecting and understanding complex diseases beyond the main effects. \cite{HUNTER, HUTTER} Significant amount of efforts have been made to conducting analysis for the investigation of the associations between disease phenotypes and interaction effects marginally, especially in GWAS. \cite{MUKH} As the disease etiology and prognosis are generally attributable to the coordinated effects of multiple genetic and environment factors, as well as the G$\times$E interactions, joint analysis has provided a powerful alternative to dissect G$\times$E interactions. 

\begin{figure}[t!]
	\centering
	\includegraphics[width=0.5\textwidth]{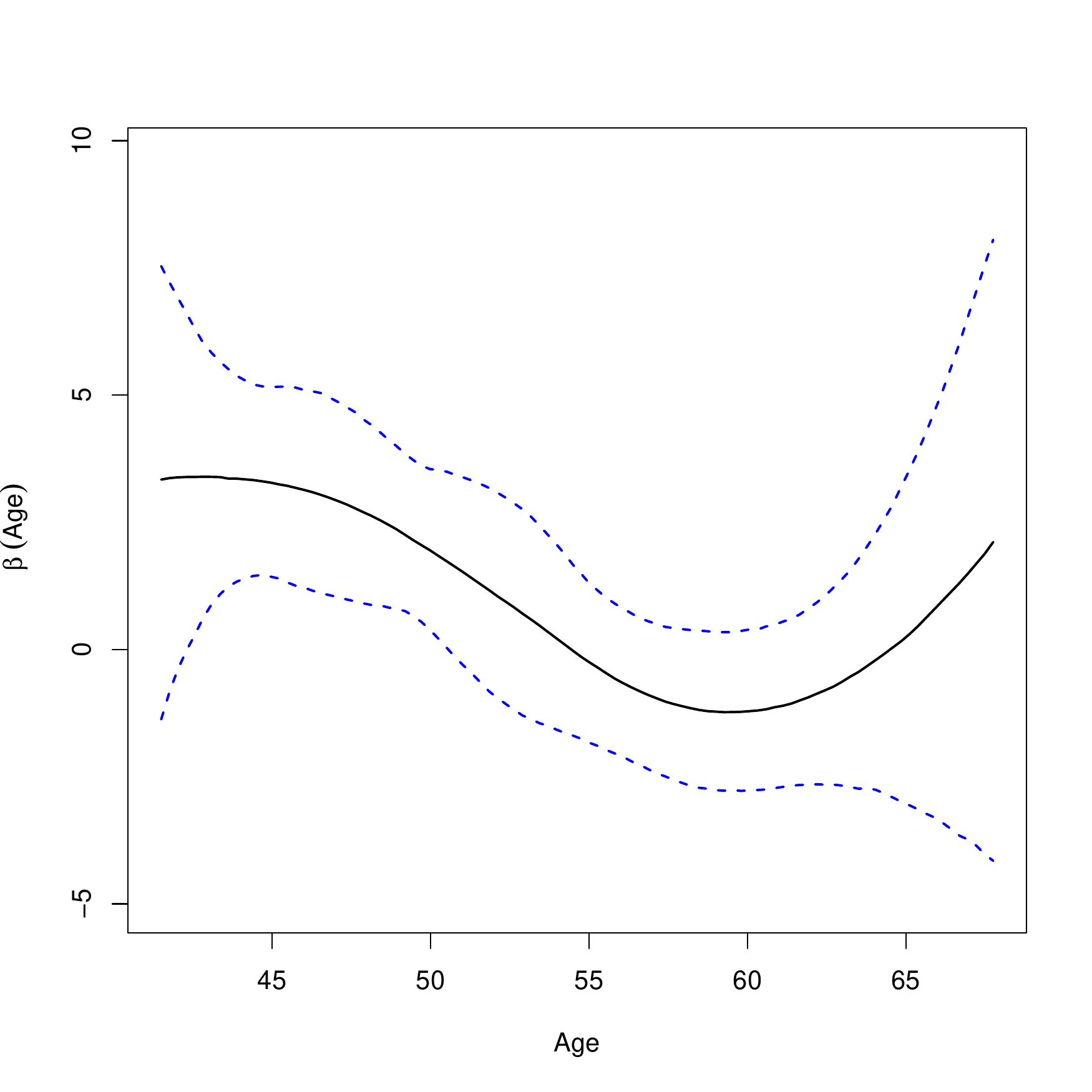}
	\caption{Non-linear G$\times$E effect of SNP rs1106380 from the Nurses' Health Study (NHS) data. The blue dashed lines represent the 95\% credible region.}
	\label{fig:curve1}
\end{figure}

From the statistical modeling perspective, the interactions can be described as the product of variables corresponding to genetic and environmental factors. With the main G and E effects, as well as their interactions, the contribution of genetic variants to disease phenotype can be expressed as a linear function of the environmental factor. Such a linear interaction assumption does not necessarily hold true in practice. Taking the Nurses' Health Study (NHS) data analyzed in this article as an example, we are interested in examining how the SNP effects on weight are mediated by age as the environmental factor. {The range of subjects' age in the NHS data is from 41 to 68. As reported, for type 2 diabetes, the average age for the onset is 45 years.\cite{CDC} Therefore, the presence of rs1106380$\times$age interaction is roughly within such a range.} We fit a Bayesian marginal model to SNP rs1106380 by using a non-parametric method to model the G$\times$E interaction while accounting for effects from clinical covariates. A 95$\%$ credible region has also been provided. Figure \ref{fig:curve1} clearly suggests that the linear interaction assumption is violated. Mis--specifying the form of interactions will lead to biased identification of important effects and inferior prediction performance. 


The non-linear G$\times$E interactions have been first conducted in marginal analysis, including Ma et al. \cite{MSJ} and Wu and Cui \cite{WU2013}. Motivated by the set based association analysis, the modeling strategy has been adopted to investigate how genetic variants in a set, such as the gene set, pathways or networks, are mediated by one or multiple types of environmental exposures to influence disease risk. The set--based modeling incorporating the nonlinear G$\times$E interactions is essentially a joint analysis
with high-dimensional covariates. Recently, penalized variable selection methods have emerged as a promising tool to capture G$\times$E interactions that might be only weak or moderate individually, but that are strong collectively.\cite{WU2014, WU2018, WMY2018,YAQI,WSC,WU2019}


Penalization methods have been first coined in Tibshirani \cite{LASSO}, which has also pointed out the connection between penalization and the corresponding Bayesian variable selection methods. In particular, the LASSO estimate can be interpreted as the posterior mode estimate when identical and independent Laplace prior has been imposed on each component of the coefficient vector under penalized least square loss. Park and Casella \cite{PARK} has further refined the prior as a conditional Laplace prior within the fully Bayesian framework to guarantee the unimodality of the posterior distribution. As LASSO belongs to the family of penalized estimate induced by the $\ell_{q}$ norm penalty with $q$=1, the Bayesian counterpart of penalization methods have been generalized to accommodate more complex data structure with other penalty functions, such as elastic net, fused LASSO and group LASSO. These extensions can also be formulated within the Bayesian framework with a similar rationale of specifying priors.\cite{KYU}


As penalization is tightly connected to Bayesian methods, the development of novel Bayesian variable selection will significantly broaden the scope of variable selection methods for G$\times$E interaction studies, which will provide us fresh perspectives and promising results not offered by the existing studies. However, our limited literature review indicates that Bayesian variable selection has not been thoroughly conducted in existing G$\times$E studies, especially for nonlinear interactions. For example, Liu et al. \cite{LIUMA} has developed a Bayesian mixture model to identify important G$\times$E and G$\times$G interaction effects through indicator model selection. Variable selection has been achieved by examining the posterior inclusion probability. Under a two-phase sampling design, Ahn et al. \cite{AHN} has considered Bayesian variable selection on G$\times$E interactions using spike--and--slab priors. Both studies cannot handle nonlinear interactions. More pertinent to the penalization, Li et al. \cite{LJH} has developed a Bayesian group LASSO for non-parametric varying coefficient models, where the non-linear interaction is expressed as a linear combinations of Legendre polynomials, and the identification of G$\times$E interactions amounts to the shrinkage selection of polynomials on the group level using multivariate Laplace priors. Li et al. \cite{LJH} has been built upon the Laplace prior adopted in Bayesian LASSO, therefore the coefficients cannot be shrunken to zero exactly in order to achieve the "real" sparsity, 

Accounting for nonlinear effects in G$\times$E studies has deeply rooted in structured variable selection for high dimensional data.\cite{ZHANGHH} An efficient selection procedure is expected to not only accurately pinpoint the form of nonlinear interactions, but also avoid modeling the main-effect-only case (corresponding to the non-zero constant effects) as nonparametric ones, since this type of misspecification may over--fit the data and result in loss of efficiency. To the best of our knowledge, automatic structure identification involving nonlinear effects has not been conducted in Bayesian G$\times$E studies. To overcome the aforementioned limitations, we develop a novel semi-parametric Bayesian variable selection method for G$\times$E interactions. We consider both linear and nonlinear interactions simultaneously. The interactions between a genetic factor and a discrete environmental factor are modeled parametrically, while the nonlinear interactions are modeled using varying coefficient functions. In particular, we conduct automatic structure identification via Bayesian regularization to separate the cases of G$\times$E interactions, main-effect-only and no genetic effects at all, which more flexibly captures the main and interaction effects. Besides, to shrink the coefficients of unimportant linear and nonlinear effects to zero exactly, we adopt the spike-and-slab priors in our model. The spike-and-slab priors have recently been shown as effective when being incorporated in Bayesian hierarchical framework for penalization methods, including the spike--and--slab LASSO \cite{VERO,TANG}, Bayesian fused LASSO \cite{ZLIN} and Bayesian sparse group LASSO \cite{XXF}. It leads to sparsity in the sense of exact 0 posterior estimates which are not available in Bayesian LASSO type of Bayesian shrinkage methods including Li et al.\cite{LJH} 

Motivated by the pressing need to conduct efficient Bayesian G$\times$E interaction studies accounting for the nonlinear interaction effects, the proposed semi-parametric model significantly advances from existing Bayesian variable selection methods for G$\times$E interactions in the following aspects. First, compared to studies that solely focus on linear \cite{LIUMA,AHN} or non-linear effects \cite{LJH}, the proposed one can accommodate both types of effects concurrently, thus more comprehensively describe the overall genetic architecture of complex diseases. Second, to the best of our knowledge, for G$\times$E interactions, automatic structure discovery has been considered in the Bayesian framework for the first time. Compared to Li et al. \cite{LJH}, one of the very few (or perhaps the only) literature in Bayesian variable selection for non-linear effects, our method is more fine tuned for the structured sparsity by distinguishing whether the genetic variants have nonlinear interaction, main effects only and no genetic effects at all, with the forms of coefficient functions being varying, non-zero constant and zero respectively. Third, borrowing strength from the spike--and--slab priors, we efficiently perform Bayesian shrinkage on the individual and group level simultaneously. In particular, with B--spline basis expansion, the identification of nonlinear interaction is equivalent to the selection of a group of basis functions. We develop an efficient MCMC algorithm for semi--parametric Bayesian hierarchical model. We show in both simulations and a case study that the exact sparsity significantly improves accuracy in identification of relevant main and interaction effects, as well as prediction. For fast computation and reproducible research, we implement the proposed and alternative methods in C++ and encapsulate them in a publicly available R package \href{https://cran.r-project.org/package=spinBayes}{spinBayes}.\cite{spinBayes}      




The rest of the article is organized as follows. In Section~\ref{makereference3.2}, we formulate the semi-parametric Bayesian variable selection model and derive a Gibbs sampler to compute the posterior estimates of the coefficients. We carry out the simulation studies to demonstrate the utility of our method in Section~\ref{makereference3.3}. A case study of Nurses' Health Study (NHS) data is conducted in Section~\ref{makereference3.4}.

\section{Data and Model Settings}
\label{makereference3.2}
\subsection{Partially linear varying coefficient model}
We denote the $i$th subject using subscript $i$. Let $(X_{i}, Y_{i}, Z_{i}, E_{i}, W_{i})$, $i=1,\ldots,n$ be independent and identically distributed random vectors. $Y_{i}$ is the response variable. $X_{i}$ is the $p$-dimensional design vector of genetic factors, and $Z_{i}$ and $E_{i}$ are the continuous and discrete environment factors, respectively. The clinical covariates are denoted by $q$-dimensional vector $W_{i}$. In the NHS data, the response variable is weight, and $X_{i}$ represents SNPs. We consider age and the indicator of history of hypertension for $Z_{i}$ and $E_{i}$, correspondingly. Height and total physical activity are used as clinical covariates, so $q$ is 2. Now consider the following partially linear varying coefficient model
\begin{equation}\label{equr:vc}
Y_{i}= \beta_{0}(Z_{i}) + \sum_{j=1}^{p}\beta_{j}(Z_{i})X_{ij} + \sum_{t=1}^{q}\alpha_{t} W_{it} + \zeta_{0} E_{i} + \sum_{j=1}^{p}\zeta_{j}E_{i}X_{ij} +\epsilon_{i}
\end{equation}
where $\beta_{j}(\cdot)$ is a smoothing varying coefficient function, $\alpha_{t}$ is the coefficient of the $t$th clinical covariates, $\zeta_{0}$ is the coefficient of the discrete E factor, and $\zeta_{j}$ is the coefficient of the interaction between the $j$th G factor $X_{j}=(X_{1j},\ldots,X_{nj})^\top $ and $E_{i}$. 
The random error $\epsilon_{i}\sim \text{N}(0, \sigma^{2})$.

Here only two environmental factors, $Z_{i}$ and $E_{i}$, are considered for the simplicity of notation. Their interactions with the G factor are modeled as non--linear and linear forms, respectively. The model can be readily extended to accommodate multiple E factors.


\subsection{Basis expansion for structure identification}
As we discussed, distinguishing the case of main-effect-only from nonlinear G$\times$E interaction is necessary since mis-specification of the effects cause over-fitting. The following basis expansion is necessary for the separation of different types of effects. We approximate the varying coefficient function $\beta_{j}(Z_{i})$ via basis expansion. Let $q_{n}$ be the number of basis functions 
\begin{equation*}\label{equr:bs}
\beta_{j}(Z_{i})\approx \sum_{k=1}^{q_{n}}\tilde{B}_{jk}(Z_{i})\tilde{\gamma}_{jk} = \tilde{B}_{j}(Z_{i})^\top  \tilde{\gamma}_{j}
\end{equation*}
where $\tilde{B}_{j}(Z_{i}) = (\tilde{B}_{j1}(Z_{i}),\ldots,\tilde{B}_{jq_{n}}(Z_{i}))^\top $ is a set of normalized B spline basis, and $\tilde{\gamma}_{j}=(\tilde{\gamma}_{j1}, \ldots, \tilde{\gamma}_{jq_{n}})^\top$ is the coefficient vector. By changing of basis, the aforementioned basis expansion is equivalent to
\begin{equation*}\label{equr:bs2}
\beta_{j}(\cdot)\approx \sum_{k=1}^{q_{n}}\tilde{B}_{jk}(\cdot)\tilde{\gamma}_{jk} \doteq \gamma_{j1} + \tilde{B}_{j*}(\cdot)^\top \gamma_{j*}
\end{equation*}
where $\tilde{B}_{j*}(Z_{i}) = (\tilde{B}_{j2}(Z_{i}),\ldots,\tilde{B}_{jq_{n}}(Z_{i}))^\top $. $\gamma_{j1}$ and $\gamma_{j*}=(\gamma_{j2}, \ldots, \gamma_{jq_{n}})^\top$ correspond to the constant and varying  components of $\beta_{j}(\cdot)$, respectively. The intercept function can be approximated similarly as  $\beta_{0}(\cdot)\approx \sum_{k=1}^{q_{n}}\tilde{B}_{0k}(\cdot)\tilde{\eta}_{k} \doteq \eta_{1} + \tilde{B}_{0*}(\cdot)^\top \eta_{*} $. Define $\gamma_{j}=(\gamma_{j1}, (\gamma_{j*})^\top )^\top $, $\eta=(\eta_{1}, (\eta_{*})^\top )^\top $, $B_{j}(Z_{i})=(1,(\tilde{B}_{j*}(Z_{i}))^\top )^\top \doteq(B_{j1}(Z_{i}),\dots,B_{jq_{n}}(Z_{i}))^\top $ and $B_{0}(Z_{i})=(1,(\tilde{B}_{0*}(Z_{i}))^\top )^\top $. Collectively, model (\ref{equr:vc}) can be rewritten as

\begin{equation*}\label{equr:vcbs}
\begin{aligned}
Y_{i} & = B_{0}(Z_{i})^\top \eta + \sum_{j=1}^{p}B_{j}(Z_{i})^\top  \gamma_{j}X_{ij}+ \sum_{t=1}^{q}\alpha_{t} W_{it} + \zeta_{0} E_{i} + \sum_{j=1}^{p}\zeta_{j}E_{i}X_{ij} +\epsilon_{i} \\
& = B_{0}(Z_{i})^\top \eta + \sum_{j=1}^{p}(X_{ij}\gamma_{j1} + U_{ij}^\top  \gamma_{j*}) +  W_{i}^\top \alpha + E_{i}^\top \zeta_{0} + T_{i}^\top \zeta +\epsilon_{i}
\end{aligned}
\end{equation*}
where $U_{ij} = (B_{j2}(Z_{i})X_{ij}, \dots, B_{jq_{n}}(Z_{i})X_{ij})^\top $, $\alpha=(\alpha_{1},\dots,\alpha_{q})^\top $, $T_{i} = (X_{i1}E_{i}, \dots, X_{ip}E_{i})^\top $, and $\zeta=(\zeta_{1},\dots,\zeta_{p})^\top$.
{Note that basis functions have been widely adopted for modeling the functional type of coefficient in general semi-parametric models, as well as functional regression analysis.\cite{KO1,KO2,ZHU} For a comprehensive review of literature in this area, please refer to Morris.\cite{MOR}}

\subsection{Semi-parametric Bayesian variable selection}

The proposed semi-parametric model is of ``large $p$, small $n$" nature. First, not all the main and interaction effects are associated with the phenotype. Second, we need to further determine for the genetic variants, whether they have nonlinear interactions, or main effect merely, or no genetic contribution to the phenotype at all. Therefore, variable selection is demanded. 

From the Bayesian perspective, variable selection falls into the following four categories: (1) indicator model selection, (2) stochastic search variable selection, (3) adaptive shrinkage and (4) model space method.\cite{HARA} Among them, adaptive shrinkage methods solicit priors based on penalized loss function, which leads to sparsity in the Bayesian shrinkage estimates. For example, within the Bayesian framework, LASSO and group LASSO estimates can be understood as the posterior mode estimates when univariate and multivariate independent and identical Laplace priors are placed on the individual and group level of regression coefficients, respectively.\cite{PARK,LJH}    

The proposed one belongs to the family of adaptive shrinkage  Bayesian variable selection. For convenience of notation, we first define the approximated least square loss function as follows:
\begin{equation*}\label{equr:loss}
\tilde{L}(\eta,\gamma,\alpha,\zeta_{0},\zeta)  = \lVert Y-B_{0}\eta-\sum_{j=1}^{p}X_{j}\gamma_{j1}-\sum_{j=1}^{p}U_{j}\gamma_{j*}-W\alpha-E\zeta_{0}-T\zeta\rVert^{2}
\end{equation*}
where $Y=(Y_{1},\ldots,Y_{n})^\top $, $B_{0}=(B_{0}(Z_{1}),\ldots,B_{0}(Z_{n}))^\top $, $U_{j}=(U_{1j},\ldots,U_{nj})^\top $, $W=(W_{1},\ldots,W_{n})^\top $ and $T=(T_{1},\ldots,T_{n})^\top $.
Let $\theta=(\eta^\top, \gamma^\top, \alpha^\top, \zeta_{0}, \zeta^\top)^\top$ be the vector of all the parameters. Then the corresponding penalized loss function is
\begin{equation}\label{equr:obj}
\tilde{L}(\eta,\gamma,\alpha,\zeta_{0},\zeta) + \lambda_{e}\sum_{j=1}^{p}\left| \zeta_{j}\right| + \lambda_{c}\sum_{j=1}^{p}\left| \gamma_{j1}\right| + \lambda_{v}\sum_{j=1}^{p}\lVert \gamma_{j*}\rVert_{2}
\end{equation}

The formulation of (\ref{equr:obj}) has been primarily driven by the need to accommodate linear and nonlinear G$\times$E interaction while avoiding mis-specification of the main-effect-only as nonlinear interactions. Here $\gamma_{j1}$ is the coefficient for the main effect of the $j$th genetic factor $X_{j}$, and the $\ell_{2}$ norm of the spline coefficients  $\lVert \gamma_{j*}\rVert_{2}$ is corresponding to the varying parts of $\beta_{j}(\cdot)$. If $\lVert \gamma_{j*}\rVert_{2}=0$, then there is no nonlinear interaction between $X_{j}$ and continuous environment factor $Z$. Furthermore, if $\gamma_{j1}=0$, then $X_{j}$ has no main effect and is not associated with the phenotype. Similarly, the linear interaction between $X_{j}$ and the discrete environment factor $E$ is determined by $\zeta_{j}$. $\zeta_{j}$=0 indicates that there is no linear interaction. Overall, the penalty functions in (\ref{equr:obj}) provide us the flexibility to achieve identification of structured sparsity through variable selection. Note that the main effects of environmental exposures $Z$ and $E$ are of low dimensionality, thus they are not subject to selection. Therefore, for the current G$\times$E interaction study, we are particular interested in conducting Bayesian variable selection on both the \textbf{individual level} of $\gamma_{j1}$ and $\zeta_{j}$ ($j=1,\dots, p$), and the \textbf{group level} of $\gamma_{j*}\, (j=1,\dots, p)$.


\textbf{Laplacian shrinkage on individual level effects.} Following the fully Bayesian analysis for LASSO proposed in Park and Casella \cite{PARK}, we impose the individual-level shrinkage on genetic main effects and linear G$\times$E interactions by adopting i.i.d. conditional Laplace prior on $\gamma_{j1}$ and $\zeta_{j}$ ($j=1,\dots, p$) 
\begin{equation}\label{equr:bl}
\begin{aligned}
\pi(\gamma_{11},\ldots,\gamma_{p1}|\sigma^{2}) &=\prod_{j=1}^{p}\frac{\lambda_{c}}{2\sigma}\exp\Big\{-\frac{\lambda_{c}}{\sigma}\left|\gamma_{j1}\right|\Big\} \\
\pi(\zeta_{1},\ldots,\zeta_{p}|\sigma^{2}) &=\prod_{j=1}^{p}\frac{\lambda_{e}}{2\sigma}\exp\Big\{-\frac{\lambda_{e}}{\sigma}\left|\zeta_{j}\right|\Big\}
\end{aligned}
\end{equation}
The above Laplace priors can be expressed as scale mixture of normals \cite{MALLO} 
\begin{equation}\label{equr:smn}
\begin{aligned}
\pi(\gamma_{j1}|\tau_{cj}^{2}, \sigma^{2}) & \stackrel{ind}{\thicksim} \text{N}(0, \, \sigma^{2}\tau_{cj}^{2})\\ 
\tau_{cj}^{2} &\stackrel{ind}{\thicksim} \frac{\lambda_{c}^{2}}{2} \exp\Big\{-\frac{\lambda_{c}^{2}}{2}\tau_{cj}^{2}\Big\}\\
\pi(\zeta_{j}|\tau_{ej}^{2},\sigma^{2}) &\stackrel{ind}{\thicksim} \text{N}(0, \, \sigma^{2}\tau_{ej}^{2})\\ 
\tau_{ej}^{2} &\stackrel{ind}{\thicksim} \frac{\lambda_{e}^{2}}{2} \exp\Big\{-\frac{\lambda_{e}^{2}}{2}\tau_{ej}^{2}\Big\}
\end{aligned}
\end{equation}
It is easy to show that, after integrating out $\tau_{cj}^{2}$ and $\tau_{ej}^{2}$, (\ref{equr:smn}) leads to the same priors in (\ref{equr:bl}).

\textbf{Laplacian shrinkage on group level effects.} Kyung et al. \cite{KYU} extended the Bayesian LASSO to a more general form that can represent the group LASSO by adopting a multivariate Laplace prior. We follow the strategy and let the prior for $\gamma_{j*}\, (j=1,\dots, p)$ be
\begin{equation}\label{equr:mlp}
\pi(\gamma_{j*}|\sigma^{2}) \propto \exp \Big\{ -\frac{\sqrt{L}\lambda_{v}}{\sigma} \lVert \gamma_{j*} \rVert_{2} \Big\}
\end{equation}
where $L=q_{n}-1$ is the size of the group, $(\frac{\sqrt{L}\lambda_{v}}{\sigma})^{-1}$ is the scale parameter of the multivariate Laplace and $\sqrt{L}$ terms adjusts the penalty for the group size. $\sqrt{L}$ can be dropped from the formula when all the groups have the same size. In this study, we use the same number of basis functions for all parameters, and thus $L$ is the same for all groups. For completeness, we still include $\sqrt{L}$ in (\ref{equr:mlp}) for possible extension to varying group sizes in the future. Similar to the (\ref{equr:smn}), this prior can be expressed as a gamma mixture of normals
\begin{equation}\label{equr:gmn}
\begin{aligned}
\pi(\gamma_{j*}|\tau_{vj}^{2},\sigma^{2}) &\stackrel{ind}{\thicksim} \text{N}_{L}(0, \, \sigma^{2}\tau_{vj}^{2} \textbf{I}_{L}) \\ 
\tau_{vj}^{2} & \stackrel{ind}{\thicksim} \text{Gamma} \Big( \frac{L+1}{2}, \, \frac{L\lambda_{v}^{2}}{2} \Big)
\end{aligned}
\end{equation}
where $\frac{L+1}{2}$ is the shape parameter and $\frac{L\lambda^{2}}{2}$ is the rate parameter of the Gamma distribution. After integrating out $\tau_{vj}^{2}$ in (\ref{equr:gmn}), the conditional prior on $\gamma_{j*}$ has the desired form in (\ref{equr:mlp}). Priors in (\ref{equr:smn}) and (\ref{equr:gmn}) can lead to a similar performance as the general LASSO model in (\ref{equr:obj}), by imposing individual shrinkage on $\gamma_{j1}$ and $\zeta_{j}$ and group level shrinkage on $\gamma_{j*}$, respectively. 

\textbf{Spike-and-slab priors on both individual and group level effects.}
Compared with (\ref{equr:obj}), priors in (\ref{equr:smn}) and (\ref{equr:gmn}) cannot shrink the posterior estimates to exact 0. Li et al. \cite{LJH} has such a limitation since multivariate Laplace priors have been imposed on the group level effects. One of the significant advancements of our study over existing Bayesian G$\times$E interaction studies, including Li et al. \cite{LJH}, is the incorporation of spike--and--slab priors to achieve sparsity.
For $\gamma_{j*}$, we have
\begin{equation}\label{equr:ssgl}
\begin{aligned}
\gamma_{j*}|\phi_{vj}, \tau_{vj}^{2}, \sigma^{2} &\stackrel{ind}{\thicksim} \phi_{vj} \text{N}_{L}(0, \, \text{diag}(\sigma^{2}\tau_{vj}^{2},\ldots, \sigma^{2}\tau_{vj}^{2})) + (1-\phi_{vj})\delta_{0}(\gamma_{j*}) \\ 
\phi_{vj}|\pi_{v} &\stackrel{ind}{\thicksim} \text{Bernoulli}(\pi_{v}) \\
\tau_{vj}^{2}|\lambda_{v}  &\stackrel{ind}{\thicksim} \text{Gamma}(\frac{L+1}{2}, \, \frac{L\lambda_{v}^{2}}{2})
\end{aligned}
\end{equation}
where $\delta_{0}(\gamma_{j*})$ denotes a point mass at $0_{L\times1}$ and $\pi_{v}\in [0,1]$. We introduce a latent binary indicator variable $\phi_{vj}$ for each group $j, (j=1,\ldots,p)$. $\phi_{vj}$ facilitates the variable selection by indicating whether or not the $j$th group is included in the final model. Specifically, when $\phi_{vj}=0$, the coefficient vector $\gamma_{j*}$ has a point mass density at zero which implies all predictors in the $j$th group are excluded from the final model. This is equivalent to concluding that the $j$th G factor $X_{j}$ does not have an interaction effect with the environment factor $Z$. On the other hand, when $\phi_{vj}=1$, the prior in (\ref{equr:ssgl}) reduces to the prior in (\ref{equr:gmn}) and induces the same behavior as Bayesian group LASSO. Thus, the coefficients in vector $\gamma_{j*}$ have non-zero values and the $j$th group is included in the final model.
Note that, after integrating out $\phi_{vj}$ and $\tau_{vj}^{2}$ in (\ref{equr:ssgl}), the marginal prior on $\gamma_{j*}$ is a mixture of a multivariate Laplace and a point mass at $0_{L\times1}$ as follows
\begin{equation}\label{equr:mprior}
\pi(\gamma_{j*}|\sigma^{2}) \thicksim \pi_{v}\, \text{M-Laplace}(0, \frac{\sigma}{\sqrt{L}\lambda_{v}}) + (1-\pi_{v})\delta_{0}(\gamma_{j*})
\end{equation}
When $\pi_{v}=1$, (\ref{equr:mprior}) is equivalent to (\ref{equr:mlp}). Fixing $\pi_{v}=0.5$ makes the prior essentially non-informative since it gives the equal prior probabilities to all sub-models. Instead of fixing $\pi_{v}$, we assign it a conjugate beta prior $\pi_{v} \thicksim \text{Beta}(r_{v}, \, w_{v})$ with fixed parameters $r_{v}$ and $w_{v}$. The value of $\lambda_{v}$ controls the shape of the slab part of (\ref{equr:mprior}) and determines the amount of shrinkage on the $\gamma_{j*}$. For computation convenience, we assign a conjugate Gamma hyperprior $\lambda_{v}^{2} \thicksim \text{Gamma}(a_{v}, \, b_{v})$ which can automatically accounts for the uncertainty in choosing $\lambda_{v}$ and ensure it is positive. We set $a_{v}$ and $b_{v}$ to small values so that the priors are essentially non-informative. 

{\textit{Remark}: The form in (\ref{equr:mprior}) shows that our prior combines the strength of the Laplacian shrinkage and the spike--and--slab prior. The Laplacian shrinkage is used as the slab part of the prior, which captures the signal in the data and provides the estimation for large effects. Compared with (\ref{equr:mlp}), the additional spike part (point mass at zero) in (\ref{equr:mprior}) shrinks the negligibly small effects to zeros and achieve the variable selection.}


Likewise, for $\gamma_{j1}$ and $\zeta_{j}$ $(j=1,\dots, p)$ corresponding to the individual level effects, the spike-and-slab priors can be written as
\begin{equation}\label{equr:ssl0}
\begin{aligned}
\gamma_{j1}|\phi_{cj}, \tau_{cj}^{2}, \sigma^{2} &\stackrel{ind}{\thicksim} \phi_{cj} \text{N}(0, \, \sigma^{2}\tau_{cj}^{2}) + (1-\phi_{cj})\delta_{0}(\gamma_{j1}) \\ 
\phi_{cj}|\pi_{c} &\stackrel{ind}{\thicksim} \text{Bernoulli}(\pi_{c}) \\
\tau_{cj}^{2}|\lambda_{c}  &\stackrel{ind}{\thicksim} \text{Gamma}(1, \, \frac{\lambda_{c}^{2}}{2})
\end{aligned}
\end{equation}
and
\begin{equation}\label{equr:ssle}
\begin{aligned}
\zeta_{j}|\phi_{ej}, \tau_{ej}^{2}, \sigma^{2} &\stackrel{ind}{\thicksim} \phi_{ej} \text{N}(0, \, \sigma^{2}\tau_{ej}^{2}) + (1-\phi_{ej})\delta_{0}(\zeta_{j}) \\ 
\phi_{ej}|\pi_{e} &\stackrel{ind}{\thicksim} \text{Bernoulli}(\pi_{e}) \\
\tau_{ej}^{2}|\lambda_{e}  &\stackrel{ind}{\thicksim} \text{Gamma}(1, \, \frac{\lambda_{e}^{2}}{2})
\end{aligned}
\end{equation}
We assign conjugate beta prior $\pi_{c} \thicksim \text{Beta}(r_{c}, \, w_{c})$ and $\pi_{e} \thicksim \text{Beta}(r_{e}, \, w_{e})$, and Gamma priors $\lambda_{c}^{2} \thicksim \text{Gamma}(a_{c}, \, b_{c})$ and $\lambda_{e}^{2} \thicksim \text{Gamma}(a_{e}, \, b_{e})$. An inverted gamma prior for $\sigma^{2}$ can maintain conjugacy. The limiting improper prior $\pi(\sigma^{2})=1/\sigma^{2} $ is another popular choice. Parameters $\eta$, $\alpha$ and $\zeta_{0}$ may be given independent flat priors.

\subsection{Gibbs sampler}
The binary indicator variables can cause an absorbing state in the MCMC algorithm which violates the convergence condition.\cite{CHIB} To avoid this problem, we integrate out the indicator variables $\phi_{c}$, $\phi_{v}$ and $\phi_{e}$ in (\ref{equr:ssgl}), (\ref{equr:ssl0}) and (\ref{equr:ssle}). We will show that, even though $\phi_{c}$, $\phi_{v}$ and $\phi_{e}$ are not part of the MCMC chain, their values still can be easily computed at every iterations. Let $\mu = E(Y)$, the joint posterior distribution of all the unknown parameters conditional on data can be expressed as
\begingroup
\allowdisplaybreaks
\begin{equation*}\label{equr:full}
\begin{aligned}
\pi (\eta, \alpha, \zeta_{0}, \gamma_{j1}, &\tau^{2}_{c}, \pi_{c}, \lambda_{c}, \gamma_{j*}, \tau^{2}_{v}, \pi_{v}, \lambda_{v}, \zeta_{j}, \pi_{e},  \lambda_{e}, \tau^{2}_{e}, \sigma^{2}|Y) \\
\propto & (\sigma^{2})^{-\frac{n}{2}} \exp\Big\{-\frac{1}{2\sigma^{2}}(Y - \mu)^\top (Y - \mu)\Big\} \\
&\times \exp\Big(-\frac{1}{2}\eta^\top \Sigma_{\eta0}^{-1}\eta\Big) \exp\Big(-\frac{1}{2}\alpha^\top \Sigma_{\alpha0}^{-1}\alpha\Big) \exp\Big(-\frac{1}{2\sigma_{\zeta_{0}}^{2}}\zeta_{0}^{2}\Big) \\
& \times \prod_{j=1}^{p}\Bigg( \pi_{v}(2\pi\sigma^{2}\tau_{vj}^{2})^{-\frac{L}{2}} \exp\Big( -\frac{1}{2\sigma^{2}\tau_{vj}^{2}}\gamma_{j*}^\top \gamma_{j*}\Big) \textbf{I}_{\{\gamma_{j*} \neq 0\}}+ (1-\pi_{v})\delta_{0}(\gamma_{j*}) \Bigg) \\
& \times (\lambda_{v}^{2})^{a_{v}-1}\exp(-b_{v}\lambda_{v}^{2} )\prod_{j=1}^{p}\Bigg(\frac{L\lambda_{v}^{2}}{2}\Bigg)^{\frac{L+1}{2}} (\tau_{vj}^{2})^{\frac{L+1}{2}-1} \exp\Bigg(-\frac{L\lambda_{v}^{2}}{2}\tau_{vj}^{2}\Bigg)\\
& \times \pi_{v}^{\gamma_{v}-1}(1-\pi_{v})^{w_{v}-1} \\
& \times \prod_{j=1}^{p}\Bigg( \pi_{c}(2\pi\sigma^{2}\tau_{cj}^{2})^{-\frac{1}{2}} \exp\Big( -\frac{1}{2\sigma^{2}\tau_{cj}^{2}}\gamma_{j1}^{2}\Big) \textbf{I}_{\{\gamma_{j1} \neq 0\}}+ (1-\pi_{c})\delta_{0}(\gamma_{j1}) \Bigg) \\
& \times (\lambda_{c}^{2})^{a_{c}-1}\exp(-b_{c}\lambda_{c}^{2} )\prod_{j=1}^{p}\frac{\lambda_{c}^{2}}{2}\exp\Bigg(-\frac{\lambda_{c}^{2}}{2}\tau_{cj}^{2}\Bigg)\\
& \times \pi_{c}^{\gamma_{c}-1}(1-\pi_{c})^{w_{c}-1} \\
& \times \prod_{j=1}^{p}\Bigg( \pi_{e}(2\pi\sigma^{2}\tau_{ej}^{2})^{-\frac{1}{2}} \exp\Big( -\frac{1}{2\sigma^{2}\tau_{ej}^{2}}\zeta_{j}^{2}\Big) \textbf{I}_{\{\zeta_{j} \neq 0\}}+ (1-\pi_{e})\delta_{0}(\zeta_{j}) \Bigg) \\
& \times (\lambda_{e}^{2})^{a_{e}-1}\exp(-b_{e}\lambda_{e}^{2} )\prod_{j=1}^{p}\frac{\lambda_{e}^{2}}{2}\exp\Bigg(-\frac{\lambda_{e}^{2}}{2}\tau_{ej}^{2}\Bigg)\\
& \times \pi_{e}^{\gamma_{e}-1}(1-\pi_{e})^{w_{e}-1} \\
& \times (\sigma^{2})^{-s-1}\exp(-\frac{h}{\sigma^{2}})
\end{aligned}
\end{equation*}
\endgroup
Let $\mu_{(-\eta)}=E(Y)-B_{0}\eta$, representing the mean effect without the contribution of $\beta_{0}(Z_{i})$. The posterior distribution of $\eta$ conditional on all other parameters can be expressed as
\begin{equation*}\label{equr:posm}
\setlength{\jot}{10pt}
\begin{aligned}
\pi(\eta|& \text{rest}) \\
& \propto \pi(\eta)\pi(y|\cdot) \\
& \propto \exp\Big(-\frac{1}{2}\eta^\top \Sigma_{\eta0}^{-1}\eta\Big)\exp\Big(-\frac{1}{2\sigma^{2}}(Y-\mu)^\top (Y-\mu)\Big) \\
& \propto \exp\Big(-\frac{1}{2}\eta^\top \Sigma_{\eta0}^{-1}\eta -\frac{1}{2\sigma^{2}}(Y-B_{0}\eta-\mu_{(-\eta)})^\top (Y-B_{0}\eta-\mu_{(-\eta)})\Big) \\
& \propto \exp\Big(\eta^\top  (\Sigma_{\eta0}^{-1}+ \frac{1}{\sigma^{2}}B_{0}^\top B_{0} )\eta -\frac{2}{\sigma^{2}}(Y-\mu_{(-\eta)})^\top B_{0}\eta\Big)
\end{aligned}
\end{equation*}
Hence, the full conditional distribution of $m$ is multivariate normal $N(\mu_{\eta}, \Sigma_{\eta})$ with mean
\begin{equation*}\label{equr:meanm}
\mu_{\eta}=\Big(\Sigma_{\eta0}^{-1}+ \frac{1}{\sigma^{2}}B_{0}^\top B_{0}\Big)^{-1} \Big(\frac{1}{\sigma^{2}}(Y-\mu_{(-\eta)})^\top B_{0}\Big)^\top 
\end{equation*}
and variance
\begin{equation*}\label{equr:varm}
\Sigma_{\eta}=\Big(\Sigma_{\eta0}^{-1}+ \frac{1}{\sigma^{2}}B_{0}^\top B_{0}\Big)^{-1}
\end{equation*}
The full conditional distribution of $\alpha$ and  $\zeta_{0}$ can be obtained in similar way. 
\begin{equation*}\label{equr:palpha}
\alpha|\text{rest} \thicksim \text{N}_{q}(\mu_{\alpha},\, \Sigma_{\alpha})
\end{equation*}
where $\mu_{\alpha}=\Sigma_{\alpha} (\frac{1}{\sigma^{2}}(Y-\mu_{(-\alpha)})^\top W)^\top $ and $\Sigma_{\alpha}=(\Sigma_{\alpha0}^{-1}+ \frac{1}{\sigma^{2}}W^\top W)^{-1}$
\begin{equation*}\label{equr:malpha0}
\zeta_{0}|\text{rest} \thicksim N(\mu_{\zeta_{0}}, \Sigma_{\zeta_{0}})
\end{equation*}
where $\mu_{\zeta_{0}}=\Sigma_{\zeta_{0}} (\frac{1}{\sigma^{2}}(Y-\mu_{(-\zeta_{0})})^\top E)$ and $\Sigma_{\zeta_{0}}=(1/\sigma_{\zeta_{0}}^{2} + \frac{\sum_{i=1}^{n}E_{i}^{2}}{\sigma^{2}})^{-1}$. 

Denote $\mu_{(-\gamma_{j*})}=E(Y)-U_{j}\gamma_{j*}$ and $l_{vj}= \pi(\gamma_{j*} \neq 0 | \text{rest})$, the conditional posterior distribution of $\gamma_{j*}$ is a multivariate spike-and-slab distribution:

\begin{equation}\label{equr:prstar}
\gamma_{j*}|\text{rest} \thicksim l_{vj} \text{N}(\mu_{\gamma_{j*}}, \, \sigma^{2}\Sigma_{\gamma_{j*}}) + (1-l_{vj})\delta_{0}(\gamma_{j*})
\end{equation}
where $\mu_{\gamma_{j*}}=\Sigma_{\gamma_{j*}}U_{j}^\top (Y - \mu_{(-\gamma_{j*})})$ and $\Sigma_{\gamma_{j*}} = (U_{j}^\top U_{j} + \frac{1}{\tau_{vj}^{2}}\textbf{I}_{L})^{-1}$. It is easy to compute that $l_{vj}$ is equal to
\begin{equation*}\label{equr:lstar}
\setlength{\jot}{10pt}
\begin{aligned}
l_{vj} = \frac{\pi_{v}}{\pi_{v} + (1-\pi_{v})(\tau_{vj}^{2})^{\frac{L}{2}}|\Sigma_{\gamma_{j*}}|^{-\frac{1}{2}} \exp\Big(- \frac{1}{2\sigma^{2}} \lVert\Sigma_{\gamma_{j*}}^{\frac{1}{2}}U_{j}^\top(Y-\mu_{(-\gamma_{j*})})\rVert_{2}^{2} \Big)} 
\end{aligned}
\end{equation*}
The posterior distribution (\ref{equr:prstar}) is a mixture of a multivariate normal and a point mass at $0$. Specifically, at the $g$th iteration of MCMC, $\gamma_{j*}^{(g)}$ is drawn from $N(\mu_{\gamma_{j*}}, \, \Sigma_{\gamma_{j*}})$ with probability $l_{vj}$ and is set to $0$ with probability $1-l_{vj}$. If $\gamma_{j*}^{(g)}$ is set to $0$, we have $\phi_{vj}^{(g)}=0$. Otherwise $\phi_{vj}^{(g)}=1$.

Likewise, the conditional posterior distributions of $\gamma_{j1}$ and $\zeta_{j}$ are also spike-and-slab distributions. Let $\mu_{\gamma_{j1}}=\Sigma_{\gamma_{j1}}X_{j}^\top(Y-\mu_{(-\gamma_{j1})}) $ and $\Sigma_{\gamma_{j1}} = (X_{j}^\top X_{j} + \frac{1}{\tau_{cj}^{2}})^{-1}$, the full conditional distribution of $\gamma_{j1}$ is
\begin{equation*}
\gamma_{j1}|\text{rest} \thicksim l_{cj} \text{N}(\mu_{\gamma_{j1}}, \, \sigma^{2}\Sigma_{\gamma_{j1}}) + (1-l_{cj})\delta_{0}(\gamma_{j1})
\end{equation*}
where
\begin{equation*}\label{equr:l0}
\setlength{\jot}{10pt}
\begin{aligned}
l_{cj} &= \pi(\gamma_{j1} \neq 0 | \text{rest}) \\
&= \frac{\pi_{c}}{\pi_{c} + (1-\pi_{c})(\tau_{cj}^{2})^{\frac{1}{2}}(\Sigma_{\gamma_{j1}})^{-\frac{1}{2}} \exp\Big(- \frac{1}{2\sigma^{2}}\Sigma_{\gamma_{j1}} \lVert(Y-\mu_{(-\gamma_{j1})})^\top X_{j}\rVert_{2}^{2} \Big)} 
\end{aligned}
\end{equation*}
Let $\mu_{\zeta_{j}}=\Sigma_{\zeta_{j}}(Y - \mu_{(-\zeta_{j})})^\top T_{j}$ and $\Sigma_{\zeta_{j}} = (T_{j}^\top T_{j} + \frac{1}{\tau_{ej}^{2}})^{-1}$, the full conditional distribution of $\zeta_{j}$ is
\begin{equation*}
\zeta_{j}|\text{rest} \thicksim l_{ej} \text{N}(\mu_{\zeta_{j}}, \, \sigma^{2}\Sigma_{\zeta_{j}}) + (1-l_{ej})\delta_{0}(\zeta_{j})
\end{equation*}
where 
\begin{equation*}\label{equr:le}
\setlength{\jot}{10pt}
\begin{aligned}
l_{ej} &= \pi(\zeta_{j} \neq 0 | \text{rest}) \\
&= \frac{\pi_{e}}{\pi_{e} + (1-\pi_{e})(\tau_{ej}^{2})^{\frac{1}{2}}(\Sigma_{\zeta_{j}})^{-\frac{1}{2}} \exp\Big(- \frac{1}{2\sigma^{2}}\Sigma_{\zeta_{j}} \lVert(Y-\mu_{(-\zeta_{j})})^\top T_{j}\rVert_{2}^{2} \Big)} 
\end{aligned}
\end{equation*}
At the $g$th iteration, the values of $\phi_{cj}^{(g)}$ and $\phi_{ej}^{(g)}$ can be determined by whether the $\gamma_{j1}^{(g)}$ and $\zeta_{j}^{(g)}$ are set to $0$ or not, respectively. We list the conditional posterior distributions of other unknown parameters here. The details can be found in the Appendix (Section~\ref{makereferenceA.7}).
\begin{equation*}\label{ptaus2}
(\tau_{vj}^{2})^{-1}|\text{rest} \thicksim \begin{cases}
\scalebox{1}{Inverse-Gamma($\frac{L+1}{2}$,\, $\frac{L\lambda_{v}^{2}}{2}$)}& { \text{if} \; \gamma_{j*} = 0} \\[5pt]
\scalebox{1}{Inverse-Gaussian($L\lambda_{v}^{2}$, $\sqrt{\frac{L\lambda_{v}^{2}\sigma^{2}}{\lVert\gamma_{j*}\rVert_{2}^{2}}}$)}& { \text{if} \; \gamma_{j*} \neq 0}
\end{cases}
\end{equation*}

\begin{equation*}\label{equr:ptau02}
(\tau_{cj}^{2})^{-1}|\text{rest} \thicksim \begin{cases}
\scalebox{1}{Inverse-Gamma($1$,\, $\frac{\lambda_{c}^{2}}{2}$)}& { \text{if} \; \gamma_{j1} = 0} \\[5pt]
\scalebox{1}{Inverse-Gaussian($\lambda_{c}^{2}$, $\sqrt{\frac{\lambda_{c}^{2}\sigma^{2}}{\gamma_{j1}^{2}}}$)}& { \text{if} \; \gamma_{j1} \neq 0}
\end{cases}
\end{equation*}

\begin{equation*}\label{equr:ptaue2}
(\tau_{ej}^{2})^{-1}|\text{rest} \thicksim \begin{cases}
\scalebox{1}{Inverse-Gamma($1$,\, $\frac{\lambda_{e}^{2}}{2}$)}& { \text{if} \; \zeta_{j} = 0} \\[5pt]
\scalebox{1}{Inverse-Gaussian($\lambda_{e}^{2}$, $\sqrt{\frac{\lambda_{e}^{2}\sigma^{2}}{\zeta_{j}^{2}}}$)}& { \text{if} \; \zeta_{j} \neq 0}
\end{cases}
\end{equation*}
$\lambda_{v}^{2}$, $\lambda_{c}^{2}$ and $\lambda_{e}^{2}$ all have inverse-gamma posterior distributions
\begin{equation*}\label{equr:plambda}
\setlength{\jot}{3pt}
\begin{aligned}
\lambda_{v}^{2}|\text{rest} &\thicksim \text{Inverse-Gamma}(a_{v}+\frac{p(L+1)}{2}, \, b_{v}+\frac{L\sum_{j=1}^{p}\tau_{vj}^{2}}{2}) \\
\lambda_{c}^{2}|\text{rest} &\thicksim \text{Inverse-Gamma}(a_{c}+p,\, b_{c}+\frac{\sum_{j=1}^{p}\tau_{cj}^{2}}{2}) \\
\lambda_{e}^{2}|\text{rest} &\thicksim \text{Inverse-Gamma}(a_{e}+p,\, b_{e}+\frac{\sum_{j=1}^{p}\tau_{ej}^{2}}{2})  \\
\end{aligned}
\end{equation*}
$\pi_{v}$, $\pi_{c}$ and $\pi_{e}$ have beta posterior distributions
\begin{equation*}\label{equr:pphi}
\setlength{\jot}{3pt}
\begin{aligned}
\pi_{v}|\text{rest} &\thicksim \text{Beta}(r_{v}+\sum_{j=1}^{p}\textbf{I}_{\{\gamma_{j*} = 0\}}, \, w_{v}+\sum_{j=1}^{p}\textbf{I}_{\{\gamma_{j*} \neq 0\}}) \\
\pi_{c}|\text{rest} &\thicksim \text{Beta}(r_{c}+\sum_{j=1}^{p}\textbf{I}_{\{\gamma_{j1} = 0\}}, \, w_{c}+\sum_{j=1}^{p}\textbf{I}_{\{\gamma_{j1} \neq 0\}}) \\
\pi_{e}|\text{rest} &\thicksim \text{Beta}(r_{e}+\sum_{j=1}^{p}\textbf{I}_{\{\zeta_{j} = 0\}}, \, w_{e}+\sum_{j=1}^{p}\textbf{I}_{\{\zeta_{j} \neq 0\}} )  \\
\end{aligned}
\end{equation*}
Last, the full conditional distribution for $\sigma^{2}$
the posterior distribution for $\sigma^{2}$ is Inverse-Gamma($\mu_{\sigma^{2}}$, $\Sigma_{\sigma^{2}}$) where
\begin{equation*}\label{equr:psigma}
\sigma^{2}|\text{rest} \thicksim \text{Inverse-Gamma}(\mu_{\sigma^{2}}, \, \Sigma_{\sigma^{2}})
\end{equation*}
with mean
\begin{equation*}
\mu_{\sigma^{2}} = s+\frac{n+\sum\textbf{I}_{\{\gamma_{j1} \neq 0\}} +L\sum\textbf{I}_{\{\gamma_{j*}\neq0\}}+\sum\textbf{I}_{\{\zeta_{j} \neq 0\}}}{2}
\end{equation*}
and variance
\begin{equation*}
\Sigma_{\sigma^{2}} = h + \frac{(Y-\mu)^\top (Y-\mu) + \sum_{j=1}^{p}\Big((\tau_{cj}^{2})^{-1}\gamma_{j1}^{2} + (\tau_{vj}^{2})^{-1}\gamma_{j*}^\top \gamma_{j*} + (\tau_{ej}^{2})^{-1}\zeta_{j}^{2}\Big)}{2}
\end{equation*}
Under our priors setting, conditional posterior distributions of all unknown parameters have closed forms by conjugacy. Therefore, efficient Gibbs sampler can be used to simulate from the posterior distribution. 

To facilitate fast computation and reproducible research, we have implemented the proposed and all the alternative methods in C++ from the R package  \href{https://cran.r-project.org/package=spinBayes}{spinBayes}\cite{spinBayes} available from the corresponding author's github website. The package is pending a manual inspection and will be available at CRAN soon.


\section{Simulation}
\label{makereference3.3}
We compare the performance of the proposed method, Bayesian spike and slab variable selection with structural identification, termed as BSSVC-SI, to four alternatives termed as BSSVC, BVC-SI, BVC and BL, respectively. {BSSVC is the proposed method but without implementing structural identification. It does not distinguish the nonzero constant effect from the nonlinear effect.} Specifically, in BSSVC, coefficients of $q_{n}$ basis functions of $\beta_{j}$ are treated as one group and are subject to selection at the group level. {Comparison of BSSVC-SI with BSSVC demonstrate the importance of structural identification in the detection of interaction effects.} BVC-SI is similar to the proposed method, except that it does not adopt the spike-and-slab prior. BVC does not use the spike-and-slab prior and does not distinguish the constant and varying effects. {All these three alternative methods, BSSVC, BVC-SI and BVC, are different variations of the proposed BSSVC-SI, aiming to evaluate the strength of using the spike-and-slab prior and demonstrate the necessity of including structural identification.} The last alternative BL is the well-known Bayesian LASSO.\cite{PARK} 
BL assumes all interactions are linear. 
Details of the alternatives, including the prior and posterior distributions, are available in the Appendix (Section~\ref{makereferenceA.7} to Section~\ref{makereferenceA.10}). 

We consider four examples in our simulations. Under all four settings, the responses are generated from model (\ref{equr:vc}) with $n=500, p=100$ and $q=2$. Note that, the dimension of regression coefficients to be estimated after basis expansion is larger than the sample size ($n=500$). For example, when the number of basis function $q_{n}=5$, the effective dimension of regression coefficient is 604. In each example, we assess the performance in terms of identification, estimation, and prediction accuracy. We use the integrated mean squared error (IMSE) to evaluate estimation accuracy on the nonlinear effects. Let $\hat{\beta}_{j}(z)$ be the estimate of a nonparametric function $\beta_{j}(z)$, and $\{z_{m}\}^{n_{grid}}_{m=1}$ be the grid points where $\beta_{j}$ is assessed. The IMSE of $\hat{\beta}_{j}(z)$ is defined as IMSE $(\hat{\beta}_{j}(z)) = \frac{1}{n_{grid}}\sum_{m=1}^{n_{grid}} \Big\{\hat{\beta}_{j}(z_{m})- \beta_{j}(z_{m})\Big\}^{2}$. Note that IMSE$(\hat{\beta}_{j})$ reduces to MSE$(\hat{\beta}_{j})$ when $\beta_{j}$ is a constant. Identification accuracy is assessed by the number of true/false positives. Prediction performance is evaluated using the mean prediction errors on an independently generated testing dataset under the same settings.

\noindent{\textit{Example 1}}

We first generate a $n\times p$ matrix of gene expressions, where $n=500$ and $p=100$, from a multivariate normal distribution with zero mean vector. We consider an auto-regression (AR) correlation structure for gene expression data, in which gene $j$ and $k$ have correlation coefficient $\rho^{|j-k|}$, with $\rho=0.5$. For each observation, we simulate two clinical covariates 
from a multivariate normal distribution with $\rho=0.5$. The continuous and discrete environment factors $Z_{i}$ and $E_{i}$ are simulated from a Unif[0, 1] distribution and a binomial distribution, respectively. The random error $\epsilon\thicksim N(0, 1)$.

The coefficients are set as $\mu(z) = 2\sin(2\pi z)$, $\beta_{1}(z) = 2\exp(2z-1)$, $\beta_{2}(z) = -6z(1-z)$, $\beta_{3}(z) = -4z^3$, $\beta_{4}(z) = 0.5$, $\beta_{5}(z) = 0.8$, $\beta_{6}(z) = -1.2$, $\beta_{7}(z) = 0.7$, $\beta_{8}(z) = -1.1$,  $\alpha_{1} = -0.5$, $\alpha_{2} = 1$, $\zeta_{0} = 1.5$, $\zeta_{1} = 0.6$, $\zeta_{2} = 1.5$, $\zeta_{3} = -1.3$, $\zeta_{4} = 1$, $\zeta_{5} = -0.8$. We set all the rest of the coefficients to 0.

\noindent{\textit{Example 2}}

We examine whether the proposed method demonstrates superior performance over the alternatives on simulated single-nucleotide polymorphism (SNP) data. The SNP genotype data $X_{i}$ are simulated by dichotomizing expression values of each gene at the 1st and 3rd quartiles, with the 3--level (2,1,0) for genotypes (AA,Aa,aa) respectively, where the gene expression values are generated from Example 1.

\noindent{\textit{Example 3}}

In the third example, we consider a different scheme to simulate SNP data. The SNP genotype data are simulated based on a pairwise linkage disequilibrium (LD) structure. For the two minor alleles A and B of two adjacent SNPs, let $q_{1}$ and $q_{2}$ be the minor allele frequencies (MAFs), respectively. 
The frequencies of four haplotypes are calculated as $p_{AB} = q_{1}q_{2} + \delta$, $p_{ab} = (1-q_{1})(1-q_{2})+\delta$, $p_{Ab} = q_{1}(1-q_{2})-\delta$, and $p_{aB} = (1-q_{1})q_{2}-\delta$, where $\delta$ denotes the LD. 
Under Hardy-Weinberg equilibrium, SNP genotype (AA, Aa, aa) at locus 1 can be generated from a multinomial distribution with frequencies $(q^{2}_{1}, 2q_{1}(1-q_{1}),(1-q_{1})^{2})$. Based on the conditional genotype probability matrix \cite{CYH}, we can simulate the genotypes for locus 2. With MAFs 0.3 and pairwise correlation $r = 0.6$, we have $\delta=r\sqrt{q_{1}(1-q_{1})q_{2}(1-q_{2})}$.

\noindent{\textit{Example 4}}

In the last example, we consider more realistic correlation structures. Specifically, we use the real data analyzed in the next section. To reduce the computational cost, we use the first 100 SNPs from the case study. For each simulation replicate, we randomly sample 500 subjects from the dataset. The same coefficients and error distribution are adopted.

Posterior samples are collected from a Gibbs Sampler running 10,000 iterations in which the first 5,000 are burn-ins. The Bayesian estimates are the posterior medians. To estimate the prediction errors, we compute the mean squared error in 100 simulations. For both BSSVC-SI and BSSVC, we consider the median probability model (MPM) \cite{XXF,BAR} to identify predictors that are significantly associated with the response variable. Suppose we collect G posterior samples from MCMC after burn-ins. The $j$th predictor is included in the regression model at the $g$th MCMC iterations if the indicator $\phi_{j}^{(g)}=1$. Thus, the posterior probability of including the $j$th predictor in the final model is defined as
\begin{equation}\label{equr:pj}
p_{j}=\hat{\pi}(\phi_{j}=1|y)=\frac{1}{G}\sum_{g=1}^{G} \phi_{j}^{(g)}, \quad j=1,\dots, p
\end{equation}

A higher posterior inclusion probability $p_{j}$ can be interpreted as a stronger empirical evidence that the $j$th predictor has a non-zero  coefficient and therefore is associated with the response variable. The MPM model is defined as the model consisting of predictors that have posterior inclusion probability at least $\frac{1}{2}$. When the goal is to select a single model, Barbieri and Berger \cite{BAR} recommends using MPM  due to its optimal prediction performance. 

\begin{table} [h!]
	\def\arraystretch{1.5}
	\begin{center}
		\caption[Identification in simulation.]{Simulation results. $(n,p,q)$ = (500, 100, 2). mean(sd) of true positives (TP) and false positives (FP) based on 100 replicates.}\label{id1.3}
		\centering
		\fontsize{10}{12}\selectfont{
			\begin{tabu} to \textwidth{ X[1.1c] X[0.3c] X[c] X[c] X[c] X[c] X[c] X[c]}
				\hline
				&&  \multicolumn{3}{c}{BSSVC-SI}  & \multicolumn{3}{c}{BSSVC}  \\
				\cmidrule(lr){3-5} \cmidrule(lr){6-8}
				&&  Varying  & Constant & Nonzero  & Varying    & Constant & Nonzero      \\
				\textbf{Example 1} &TP	&3.00(0.00)	&4.93(0.25)	&5.00(0.00)	&3.00(0.00)	&0.00(0.00)	&5.00(0.00)										
				\\
				&FP	&0.20(0.41)	&0.00(0.00)	&0.00(0.00)	&5.00(0.26)	&0.00(0.00)	&0.10(0.31)				 			 		
				\\
				\textbf{Example 2}  &TP	&3.00(0.00)	&5.00(0.00)	&5.00(0.00)	&3.00(0.00)	&0.00(0.00)	&5.00(0.00)				
				\\
				&FP	&0.20(0.41)	&0.00(0.00)	&0.03(0.18)	&5.00(0.26)	&0.00(0.00)	&0.03(0.18)				
				\\
				\textbf{Example 3} &TP &3.00(0.00)	&4.97(0.18)	&5.00(0.00)	&3.00(0.00)	&0.00(0.00)	&5.00(0.00)		
				\\
				&FP	&0.03(0.18)	&0.07(0.37)	&0.00(0.00)	&5.03(0.18)	&0.00(0.00)	&0.10(0.31)				 
				\\
				\textbf{Example 4} &TP	&3.00(0.00)	&4.97(0.18)	&5.00(0.00)	&3.00(0.00)	&0.00(0.00)	&5.00(0.00)				
				\\
				&FP	&0.17(0.38)	&0.03(0.18)	&0.00(0.00)	&5.10(0.31)	&0.00(0.00)	&0.13(0.35)
				\\
				\hline
		\end{tabu} }
	\end{center}
	\centering
\end{table}

Table \ref{id1.3} summarized the results on model selection accuracy. The identification performance for the varying and nonzero constant effects corresponding to the continuous environment factor, and nonzero effect (linear interaction) corresponding to the discrete environment factor are evaluated separately. We can observe that the proposed model has superior performance over BSSVC. BSSVC fails to identify any nonzero constant effect and has high false positive for identifying varying effect since it lacks structural identification to separate main-effect-only case from the varying effects. On the other hand, BSSVC-SI identifies most of the true effects with very lower false positives. For example, considering the MPM in Example 1, BSSVC-SI identifies all 3 true varying effects in every iteration, with a small number of false positives 0.20(sd 0.41). It also identifies 4.93(sd 0.25) out of the 5 true constant effects without false positives. Besides, all the 5 true nonzero effects are identified without any false positives. We demonstrate the sensitivity of BSSVC-SI for variable selection to the choice of the hyper-parameters for $\pi_{v}$, $\pi_{c}$ and $\pi_{e}$ {and the the choice of the hyper-parameters for $\lambda_{v}$, $\lambda_{c}$ and $\lambda_{e}$} in the Appendix. The results are tabulated in Table \ref{sens} and Table \ref{sens3}, respectively. Both tables show that the MPM model is insensitive to different specification of the hyper-parameters.
The alternatives BVC-SI and BVC are not included here due to the lack of variable selection property. Li et al. \cite{LJH} adopts a method that is based on 95\% credible interval (95\%CI) for selecting important varying effects. In the Appendix, we show that, even adopting the 95\%CI-based selection method, the identification performance of BVC-SI and BVC are unsatisfied, especially in terms of selecting a large number of false positives (Table \ref{id1.95}).

We also examine the estimation performance. We show the results from Example 1 (Table \ref{est1}) here. The IMSE for all true varying effects, MSE for constant and nonzero effects, as well as the total squared errors for all coefficient estimates and prediction errors are provided in the Table. We observe that, across all the settings, the proposed method has the smallest prediction errors and total squared errors of coefficients estimates than all alternatives. For example, in Table \ref{est1}, the BSSVC-SI has the smallest total squared errors 0.268(sd 0.080) and prediction error 1.159(0.066) among all the approaches. The key of the superior performance lies in (1) accurate modeling of different types of main and interaction effects, and (2) the spike and slab priors for achieving sparsity. Compared with BVC-SI which has (1) but does not spike and slab prior, BSSVC-SI performs better when estimating both varying and constant coefficients. For example, the IMSE and MSE on $\beta_{0}(Z)$ and $\alpha_{1}$ are 0.049 (sd 0.017) and 0.004 (sd 0.004), respectively. While BVC-SI yields 0.067(sd 0.030) and 0.008(0.010), correspondingly. Besides, compared with BSSVC which adopts the spike and slab priors without considering structured Bayesian variable selection, BSSVC-SI has comparable estimation performance on coefficients even though BSSVC overfits the data. In addition, similar patterns have been observed in Table \ref{est2}, Table \ref{est3} and Table \ref{est4} for Examples 2, 3 and 4 respectively, in the Appendix.

{As a demonstrating example, Figure \ref{fig:curve} shows the estimated varying coefficients of the proposed model for the gene expression data in Example 1. Results from the proposed method fit the underlying trend of varying effects reasonably well. }
{Following Li et al. \cite{LJH}, we assess the convergence of the MCMC chains by the potential scale reduction factor (PSRF).\cite{PSRF, PSRF2} PSRF values close to 1 indicate that chains converge to the stationary distribution. Gelman et al. \cite{BDA} recommend using PSRF$\leq1.1$ as the cutoff for convergence, which has been adopted in our study. We compute the PSRF for each parameter and find all chains converge after the burn-ins. For the purpose of demonstration, Figure \ref{fig:cvg} shows the pattern of PSRF after burn-ins for each parameter in Figure \ref{fig:curve}. The figure clearly shows the convergence of the proposed Gibbs sampler. }

We conduct sensitivity analysis on how the smoothness specification of the parameters in the B spline affects variable selection. The results summarized in Table \ref{sens2} in the Appendix shows that the proposed model is insensitive to the smoothness specification as long as the choices on number of spline basis are sensible. In simulation, we set the degree of B spline basis $O=2$ and the number of interior knots $K=2$, which makes $q_{n}=5$.

{Computation feasibility is an important practical consideration for high-dimensional Bayesian variable selection methods. We examine the computational cost of the proposed method for finishing 10,000 MCMC iterations under different combinations of sample sizes and SNP numbers. We focus on SNP numbers since the increase is computationally more challenging than that of the covariate numbers due to basis expansion. The results summarized in Table \ref{cost} show that the proposed method is highly computationally efficient. For example, when sample size $n=1500$ and the number of gene $p=300$, the CPU time for 10,000 iterations is approximately 121 seconds. Please note that the number of regression coefficients to be estimated after basis expansion is on the order of $q_{n}p + p$, where $q_{n}$ is the number of basis functions. The term $q_{n}p$ gives the number of spline coefficients of nonlinear G$\times$E interactions and $p$ is the number of linear G$\times$E interactions.
In this example, the number of regression coefficients to be estimated is approximately 1800, higher than the sample size $n=1500$. The efficient C++ implementation of the Gibbs sampler is an important guarantee for the computational scalability. The proposed method can be potentially applied to larger datasets with a reasonable computation time.}

\begin{table} [t!]
	\def\arraystretch{1.2}
	\begin{center}
		\caption[Simulation results in Example 1.]{Simulation results in Example 1. Gene expression data $(n,p,q)$ = (500, 100, 2). mean(sd) of the integrated mean squared error (IMSE), mean squared error (MSE), total squared errors for all estimates and prediction errors based on 100 replicates.}\label{est1}
		\centering
		\fontsize{10}{12}\selectfont{
			\begin{tabu} to \textwidth{ X[c] X[c] X[c] X[c]  X[c] X[c] }
				\hline
				&  BSSVC-SI  & BSSVC  & BVC-SI & BVC & BL \\
				\hline
				\textbf{IMSE} & & & &   &   \\
				$\beta_{0}(Z)$  &0.049(0.017)	&0.050(0.017)	&0.067(0.030)	&0.066(0.028)	&0.806(0.039)								
				\\
				$\beta_{1}(Z)$  &0.052(0.028)	&0.027(0.019)	&0.090(0.051)	&0.107(0.051)	&0.139(0.060)							
				\\
				$\beta_{2}(Z)$ &0.035(0.020)	&0.026(0.014)	&0.045(0.023)	&0.050(0.021)	&0.252(0.049)							
				\\
				$\beta_{3}(Z)$  &0.033(0.025)	&0.024(0.019)	&0.081(0.057)	&0.106(0.062) &0.256(0.062)									
				\\[0.1cm]
				\textbf{MSE} & & & &   &   \\
				$\alpha_{1}$  &0.004(0.004)	&0.004(0.005)	&0.008(0.010)	&0.008(0.011)	&0.012(0.015)								
				\\
				$\alpha_{2}$  &0.004(0.005)	&0.004(269)	&0.009(0.013)	&0.009(0.013)	&0.011(0.012)							
				\\
				$\zeta_{0}$ &0.033(0.025)	&0.024(0.019)	&0.081(0.057)	&0.106(0.062)	&0.032(0.045)		
				\\[0.1cm]
				$\zeta_{1}$ &0.004(0.005)	&0.003(0.004)	&0.007(0.008)	&0.006(0.007)	&0.026(0.043)										
				\\
				$\zeta_{2}$  &0.011(0.014)	&0.009(0.011)	&0.017(0.016)	&0.017(0.016)	&0.055(0.067)										
				\\
				$\zeta_{3}$ &0.008(0.011)	&0.008(0.010)	&0.017(0.024)	&0.017(0.022)	&0.055(0.052)									
				\\
				$\zeta_{4}$ &0.014(0.017)	&0.019(0.028)	&0.020(0.025)	&0.020(0.023)	&0.042(0.052)									
				\\
				$\zeta_{5}$  &0.009(0.013)	&0.010(0.016)	&0.020(0.030)	&0.024(0.032)	&0.048(0.052)									
				\\[0.1cm]
				\textbf{Total}  &0.268(0.080)	&0.304(0.132)	&2.181(0.373)	&2.119(0.363)	&4.916(0.564)									
				\\[0.1cm]
				\textbf{Pred.} & & & &   &   \\
				\textbf{Error} &1.159(0.066)	&1.167(0.067)	&2.112(0.175)	&2.075(0.170)	&9.417(0.914)							
				\\[0.1cm]
				\hline
		\end{tabu} }
	\end{center}
	\centering
\end{table}

\section{Real Data Analysis}
\label{makereference3.4}
We analyze the data from Nurses' Health Study (NHS). We use weight as the response and focus on SNPs on chromosome 10. We consider two environment factors. The first is age which is continuous and is known to be related to the variations in the obesity level. The second is the binary indicator of whether an individual has a history of hypertension (hbp), which is a sensible candidate for a discrete environment factor. In addition, we consider two clinical covariates: height and total physical activity. In NHS study, about half of the subjects are diagnosed of type 2 diabetes (T2D) and the other half are controls without the disease. We only use health subjects in this study. After cleaning the data through matching phenotypes and genotypes, removing SNPs with minor allele frequency (MAF) less than 0.05 or deviation from Hardy--Weinberg equilibrium, the working dataset contains 1716 subjects with 35099 SNPs.

For computational convenience prescreening can be conducted to reduce the feature space to a more attainable size for variable selection. For example, Li et al. \cite{LJH} use the single SNP analysis to filter SNPs in a GWA study before downstream analysis. 
{In this study, we follow the procedure described in Ma et al. \cite{MSJ} and Wu and Cui \cite{WU2013} to screen SNPs. Specifically, we use three likelihood ratio tests with weight as the response variable to evaluate the penetrance effect of a variant under the environmental exposure. The three likelihood ratio tests have been developed to test whether the interaction effects are nonlinear, linear, constant or zero, respectively. The SNPs with p-values less than a certain cutoff (0.005) from any of the tests are kept.}
269 SNPs pass the screening.

We analyze the data by using the proposed method as well as BSSVC, the alternative without structural identification. As methods BVC-SI, BVC and BL show inferior performance in simulation, they are not considered in real data analysis. The proposed method identifies three SNPs with constant effects only, eleven SNPs with varying effects and sixteen SNPs with interactions with the hbp indicator. The BSSVC identifies twelve SNPs with varying effects and 10 SNPs with interactions with the hbp indicator. The Identification results for varying and constant effects are summarized in Table \ref{real1}. In this table, we can see that the three SNPs (rs11014290, rs2368945 and rs10787374) that are identified as constant effects only by BSSVC-SI are also selected by BSSVC. However, due to lack of structural identification, BSSVC identified them as SNPs with varying effects. The proposed method identifies rs1816002, a SNP located within gene ADAMTS14 as an important SNP with varying effect. ADAMTS14 is a member of ADAMTS metalloprotease family. Studies have shown that two members in the family, ADAMTS1 and ADAMTS13 are related to the development of obesity \cite{ADAM,ADAM13}, which suggests that ADAMTS14 may also have implications in obesity. The alternative method BSSCV fails to identify this important gene. The varying effect of the DIP2C gene SNP rs4880704 is identified by both BSSVC-SI and BSSVC. DIP2C (disco interacting protein 2 homolog C) has been found a potential epigenetic mark associated with obesity in children \cite{DIP2C} and plays an important role in the association between obesity and hyperuricemia.\cite{DIP2C-2} 
The identification results for nonzero effects (representing the interactions with the binary indicator of a history of hypertension (hbp)) are summarized in Table \ref{real2}. The interaction between rs593572 in gene KCNMA1 and hbp is identified by the proposed method. KCNMA1 (potassium calcium-activated channel subfamily M alpha 1) has been reported as an obesity gene that contributes to excessive accumulation of adipose tissue in obesity.\cite{KCNMA1} Interestingly, the main effect of KCNMA1 is not identified, which suggests that KCNMA1 only has effect in the hypertension patients group. This result could be partially explained by the observation of significant association between the genetic variation in the KCNMA1 and hypertension.\cite{KCNMA1-2}

\begin{table} [h!]
	\def\arraystretch{1.5}
	\begin{center}
		\caption{Identification results for varying and constant effects.}\label{real1}
		\centering
		\fontsize{10}{12}\selectfont{
			\begin{tabu} to 0.6\textwidth{ X[c] X[1.1c] X[c] X[c] X[c] }
				\hline
				&&  \multicolumn{2}{c}{BSSVC-SI}  & \multicolumn{1}{c}{BSSVC}  \\
				\cmidrule(lr){3-4} \cmidrule(lr){5-5}
				SNP & Gene &  V(Age)  & C  & V(Age)     \\
				rs11014290 &PRTFDC1 	&	&-1.864	 &Varying 							
				\\
				rs2368945 &RPL21P93 	&	&1.494	 &Varying 
				\\
				rs4880704 &DIP2C  &Varying	&  &Varying 
				\\
				rs1106380 &CACNB2 	&Varying	&  &Varying 
				\\
				rs2245456  &MALRD1 	&Varying	&  & 
				\\
				rs17775990 &OGDHL 	&Varying	&  &Varying 
				\\
				rs7922576  &ZNF365 	&Varying	&  &Varying 
				\\
				rs1816002 &ADAMTS14  &Varying	&  & 
				\\
				rs2784761 &RPL22P18  &Varying	&  &Varying 
				\\
				rs181652 &AC005871.1 	&Varying	&  & 
				\\
				rs10765108 &DOCK1  &Varying	&  & 
				\\
				rs2764375 &LINC00959 	&Varying	&  &Varying 
				\\
				rs10787374 &RPS6P15 	&	&2.020	 &Varying
				\\
				rs11006525 &MRPL50P4 	&Varying& &  
				\\
				rs1698417 &AC026884.1 	& &  &Varying
				\\
				rs7084791 &PPP1R3C 	&	 & &Varying 
				\\
				rs12354542 &BTF3P15 	&	 & &Varying 
				\\
				\hline
		\end{tabu} }
	\end{center}
	\centering
\end{table}

\begin{table} [h!]
	\def\arraystretch{1.5}
	\begin{center}
		\caption{Identification results for nonzero effect corresponds to the discrete environment effect.}\label{real2}
		\centering
		\fontsize{10}{12}\selectfont{
			\begin{tabu} to 0.5\textwidth{ X[1.5c] X[1.5c] X[c] X[c] }
				\hline
				&&  \multicolumn{1}{c}{BSSVC-SI}  & \multicolumn{1}{c}{BSSVC}  \\
				\cmidrule(lr){3-3} \cmidrule(lr){4-4}
				rs10740217 &CTNNA3 &-1.06 &-1.18 \\
				rs10787374 &RPS6P15 &-1.56 &-1.42 \\
				rs10795690 &AC044784.1 & &1.23 \\
				rs10829152 &ANKRD26 &1.29 &1.73 \\
				rs10999234 &PRKG1 & &1.97 \\
				rs11187761 &PIPSL &1.04 & \\
				rs11245023 &C10orf90 &-0.92 & \\
				rs11250578 &ADARB2 &-1.62 & \\
				rs12267702 &LYZL1 &1.30 &0.96 \\
				rs17767748 &BTRC &1.18 &1.15 \\
				rs2495763 &PAX2 &-1.33 &-1.12 \\
				rs4565799 &MCM10 &-0.84 &-0.98 \\
				rs593572 &KCNMA1 &1.70 & \\
				rs685578 &AL353149.1 & &-1.13 \\
				rs7075347 &AL357037.1 &1.00 & \\
				rs7911264 &HHEX &-1.30 & \\
				rs796945 &RNLS &1.89 & \\
				rs9419280 &LINC01168 &1.57 & \\
				rs997064 &PCDH15 &1.31 & \\
				\hline
		\end{tabu} }
	\end{center}
	\centering
\end{table}

The eleven varying coefficients of age that are identified by BSSVC-SI and the intercept are shown in Figure \ref{fig:curve2} in the Appendix. All estimates have clear curvature and cannot be appropriately approximated by a model assuming linear effects. It is difficult to objectively evaluate the selection performance with real data. The prediction performance may provide partial information on the relative performance of different methods. Following Yan and Huang \cite{YJHJ} and Li et al. \cite{LJH}, we refit the models selected by BSSVC-SI and BSSVC by Bayesian LASSO. The prediction mean squared errors (PMSE) based on the posterior median estimates are computed. The PMSEs are 90.66 and 95.21 for BSSVC-SI and BSSVC, respectively. We also compute the prediction performance of BVC-SI, BVC and BL, based on the models selected by the 95\% CI-based method. The PMSE is 106.26 for BVC-SI,  110.19 for BVC and 107.82 for BL. The proposed method outperforms all the competitors.

\section{Discussion}
\label{makereference3.5}
The importance of G$\times$E interactions in deciphering the genetic architecture of complex diseases have been increasingly recognized. A considerable amount of effort has been developed to dissect the G$\times$E interactions. In marginal analysis, statistical testing of G$\times$E interactions prevails, which spans from the classical linear model with interactions in a wide range of studies, such as case-control study, case only study and the two-stage screening study, to more sophisticated models, such as empirical Bayesian models, non- and semi-parametric models.\cite{COR} On the other hand, the joint methods, especially the penalized variable selection methods, for G$\times$E interactions, have been motivated by the success of gene set based association analysis over marginal analysis, as demonstrated in Wu and
Cui \cite{WU2013BB}, Wu et al. \cite{WU2012} and Schaid et al.\cite{SCHA}. Recently, multiple penalization methods have been proposed to identify important G$\times$E interactions under parametric, semi-parametric and non-parametric models recently.\cite{WU2014, WU2018, WSC, WU2019}

Within the Bayesian framework, non-linear interaction has not been sufficiently considered for G$\times$E interactions. Furthermore, incorporation of the structured identification to determine whether the genetic variants have non-linear interaction, or main-effect-only, or no genetic influences at all is particularly challenging. In this study, we have proposed a novel semi-parametric Bayesian variable selection method to simultaneously pinpoint important G$\times$E interactions in both linear and nonlinear forms while conducting automatic structure discovery. We approximate the nonlinear interaction effects using B splines, and develop a Bayesian hierarchical model to accommodate the selection of linear and nonlinear G$\times$E interactions. For the nonlinear effects, we achieve the separation of varying, non-zero constant and zero coefficient functions through changing of spline basis, corresponding to cases of G$\times$E interactions, main effects only (no G$\times$E interactions) and no genetic effects. This automatic separation of different effects, together with the identification of linear interaction, lead to selection of important coefficients on both individual and group levels. Within our Bayesian hierarchical model, the group and individual level shrinkage are induced through assigning spike-and-slab priors with the slab parts coming from a multivariate Laplace distribution on the group of spline coefficients and univariate Laplace distribution on the individual coefficient, correspondingly. We have developed an efficient Gibbs sampler and implemented in R with core modules developed in C++, which guarantees fast computation in MCMC estimation. The superior performance of the proposed method over multiple alternatives has been demonstrated through extensive simulation studies and a case study.

The cumulative evidence has indicated the effectiveness of penalized variable selection methods to pinpoint important G$\times$E interactions. Bayesian variable selection methods, however, have not been widely adopted in existing G$\times$E studies. The proposed semi-parametric Bayesian variable selection method has the potential to be extended to accommodate a diversity forms of complex interaction structures under the varying index coefficient models and models alike, as summarized in Ma and Song\cite{SJM}. Other possible extensions include Bayesian semi-parametric interaction analysis for integrating multiple genetic datasets.\cite{YLI} Investigations of all the aforementioned extensions are postponed to the future.   


\section*{Acknowledgments}
We thank the associate editor and reviewers for their careful review and insightful comments, which have led to a significant improvement of this article. This study has been partly supported by the National Institutes of Health (CA191383, CA204120), the VA Cooperative Studies Program of the Department of VA, Office of Research and Development, an innovative research award from KSU Johnson Cancer Research Center and a KSU Faculty Enhancement Award. Zhang's work is supported by NIAID/NIH (R01AI121226). Funding support for the GWAS of Gene and Environment Initiatives in Type 2 Diabetes was provided through the NIH Genes, Environment and Health Initiative [GEI] (U01HG004399). The datasets used for the analyses described in this manuscript were obtained from dbGaP through accession number phs000091.v2.p1.





\subsection*{Conflict of interest}

The authors declare no potential conflict of interests.

%
%
%

\bibliography{references}%

\appendix

\section{Additional simulation results}

\subsection{Hyper-parameters sensitivity analysis}
We demonstrate the sensitivity of BSSVC-SI for variable selection to the choice of the hyperparameters for $\pi_{v}$, $\pi_{c}$ and $\pi_{e}$. We consider five different Beta priors: (1) Beta(0.5, 0.5) which is a U-shape curve between $(0,1)$; (2) Beta(1, 1) which is a essentially a uniform prior; (3) Beta(2, 2) which is a quadratic curve; (4) Beta(1, 5) which is highly right-skewed; (5) Beta(5, 1) which is highly left-skewed. As a demonstrating example, we use the same setting of Example 2 to generate data. Table \ref{sens} shows the identification performance of the median thresholding model (MPM) with different Beta priors. For all choices of Beta priors, the MPM model is very stable for both the proposed model BSSVC-SI and the alternative BSSVC. Also BSSVC-SI correctly identifies almost all true effects with low false positives in all cases. {Therefore, we simply use Beta$(1,1)$ as the prior for $\pi_{v}$, $\pi_{c}$ and $\pi_{e}$ in this study.}

{We also evaluate the sensitivity of BSSVC-SI to the choice of the Gamma hyperpriors on $\lambda_{v}$, $\lambda_{c}$ and $\lambda_{e}$. We test the shape parameter of the Gamma prior for five different values: $\{0.1, 0.5, 1, 2, 5\}$. This ranges from highly skewed exponential shape to highly diffuse unimodal shape. We fix the rate parameter at $\{1, 2, 5\}$ and test different combinations of shape and rate parameters on a two-dimensional grid. In Table \ref{sens3}, we show the simulation results of some representative cases under the scenarios of Example 2. BSSVC-SI model has stable performance with high TP and low FP for different Gamma priors. Similar patterns are observed for all other cases. In this study, we use Gamma(1, 1) for $\lambda_{v}$, $\lambda_{c}$ and $\lambda_{e}$ under all scenarios.} 

\begin{table} [h!]
	\def\arraystretch{1.5}
	\begin{center}
		\caption[Sensitivity analysis on hyper-parameters.]{Sensitivity analysis. $(n,p,q)$ = (500, 100, 2). mean(sd) of true positives (TP) and false positives (FP) based on 100 replicates.}\label{sens}
		\centering
		\fontsize{10}{12}\selectfont{
			\begin{tabu} to \textwidth{ X[1.3l] X[0.3c] X[c] X[c] X[c] X[c] X[c] X[c]}
				\hline
				&&  \multicolumn{3}{c}{BSSVC-SI}  & \multicolumn{3}{c}{BSSVC}  \\
				\cmidrule(lr){3-5} \cmidrule(lr){6-8}
				&&  Varying  & Constant & Nonzero  & Varying    & Constant & Nonzero      \\
				\textbf{Beta}(0.5, 0.5) &TP	&3.00(0.00)	&5.00(0.00)	&5.00(0.00)	&3.00(0.00)	&0.00(0.00)	&5.00(0.00)				
				\\
				&FP	&0.07(0.25)	&0.00(0.00)	&0.00(0.00)	&5.07(0.25)	&0.00(0.00)	&0.03(0.18)				
				\\
				\textbf{Beta}(1, 1)  &TP	&3.00(0.00)	&5.00(0.00)	&5.00(0.00)	&3.00(0.00)	&0.00(0.00)	&5.00(0.00)				
				\\
				&FP	&0.07(0.25)	&0.00(0.00)	&0.03(0.18)	&5.00(0.00)	&0.00(0.00)	&0.10(0.31)					
				\\
				\textbf{Beta}(2, 2) &TP &3.00(0.00)	&4.93(0.25)	&5.00(0.00)	&3.00(0.00)	&0.00(0.00)	&5.00(0.00)						
				\\
				&FP	&0.20(0.48)	&0.00(0.00)	&0.00(0.00)	&4.97(0.18)	&0.00(0.00)	&0.10(0.31)				
				\\
				\textbf{Beta}(1, 5) &TP	&3.00(0.00)	&4.97(0.18)	&5.00(0.00)	&3.00(0.00)	&0.00(0.00)	&5.00(0.00)							
				\\
				&FP	&0.17(0.46)	&0.00(0.00)	&0.00(0.00)	&5.00(0.00)	&0.00(0.00)	&0.03(0.18)				
				\\
				\textbf{Beta}(5, 1) &TP	&3.00(0.00)	&5.00(0.00)	&5.00(0.00)	&3.00(0.00)	&0.00(0.00)	&5.00(0.00)							
				\\
				&FP	&0.27(0.52)	&0.07(0.25)	&0.03(0.18)	&5.07(0.25)	&0.00(0.00)	&0.27(0.58)					
				\\
				\hline
		\end{tabu} }
	\end{center}
	\centering
\end{table}

\begin{table} [h!]
	\def\arraystretch{1.5}
	\begin{center}
		\caption[Sensitivity analysis on hyper-parameters.]{{Sensitivity analysis. $(n,p,q)$ = (500, 100, 2). mean(sd) of true positives (TP) and false positives (FP) based on 100 replicates.}}\label{sens3}
		\centering
		\fontsize{10}{12}\selectfont{
			\begin{tabu} to \textwidth{ X[1.5l] X[0.3c] X[c] X[c] X[c] X[c] X[c] X[c]}
				\hline
				&&  \multicolumn{3}{c}{BSSVC-SI}  & \multicolumn{3}{c}{BSSVC}  \\
				\cmidrule(lr){3-5} \cmidrule(lr){6-8}
				&&  Varying  & Constant & Nonzero  & Varying    & Constant & Nonzero      \\
				\textbf{Gamma}(0.1, 1) &TP	&3.00(0.00)	&4.93(0.26)	&5.00(0.00)	&3.00(0.00)	&0.00(0.00)	&5.00(0.00)
				\\
				&FP	&0.20(0.41)	&0.07(0.26)	&0.00(0.00)	&5.00(0.00)	&0.00(0.00)	&0.07(0.26)											
				\\
				\textbf{Gamma}(0.5, 2)  &TP	&3.00(0.00)	&5.00(0.00)	&5.00(0.00)	&3.00(0.00)	&0.00(0.00)	&5.00(0.00)
				\\
				&FP	&0.07(0.26)	&0.00(0.00)	&0.00(0.00)	&5.00(0.00)	&0.00(0.00)	&0.00(0.00)											
				\\
				\textbf{Gamma}(1, 1) &TP &3.00(0.00)	&5.00(0.00)	&5.00(0.00)	&3.00(0.00)	&0.00(0.00)	&5.00(0.00)						
				\\
				&FP	&0.07(0.25)	&0.00(0.00)	&0.03(0.18)	&5.00(0.00)	&0.00(0.00)	&0.10(0.31)				
				\\
				\textbf{Gamma}(1, 5) &TP &3.00(0.00)	&4.93(0.26)	&5.00(0.00)	&3.00(0.00)	&0.00(0.00)	&5.00(0.00)
				\\
				&FP	&0.07(0.26)	&0.00(0.00)	&0.07(0.26)	&5.00(0.00)	&0.00(0.00)	&0.07(0.26)								
				\\
				\textbf{Gamma}(2, 5) &TP	&3.00(0.00)	&4.93(0.26)	&5.00(0.00)	&3.00(0.00)	&0.00(0.00)	&5.00(0.00)
				\\
				&FP	&0.13(0.35)	&0.07(0.26)	&0.00(0.00)	&4.93(0.26)	&0.00(0.00)	&0.00(0.00)						
				\\
				\textbf{Gamma}(5, 1) &TP	&3.00(0.00)	&5.00(0.00)	&5.00(0.00)	&3.00(0.00)	&0.00(0.00)	&5.00(0.00)
				\\
				&FP	&0.20(0.41)	&0.00(0.00)	&0.00(0.00)	&5.07(0.26)	&0.00(0.00)	&0.20(0.41)	
				\\
				\hline
		\end{tabu} }
	\end{center}
	\centering
\end{table}
\subsection{Variable selection based on 95\% credible interval}
Alternatives BVC-SI and BVC lack for the variable selection property. In order to create sparsity on the coefficients estimated by these two methods, we consider a 95\% credible interval based method used in Li et al. \cite{LJH}. Specifically, a varying effect is included in the final model if at least one of its spline coefficients has a two-sided 95\% credible interval that does not cover zero. Similarly, a constant effect is included in the final model if the two-sided 95\% credible interval of its spline coefficient does not cover zero. The same rule applies to the linear interaction effects. The results are tabulated in Table \ref{id1.95}.

\begin{table} [h!]
	\def\arraystretch{1.5}
	\begin{center}
		\caption[Identification in simulation.]{Simulation results. $(n,p,q)$ = (500, 100, 2). mean(sd) of true positives (TP) and false positives (FP) based on 100 replicates.}\label{id1.95}
		\centering
		\fontsize{10}{12}\selectfont{
			\begin{tabu} to \textwidth{ X[1.1c] X[0.3c] X[c] X[c] X[c] X[c] X[c] X[c]}
				\hline
				&&  \multicolumn{3}{c}{BVC-SI}  & \multicolumn{3}{c}{BVC}  \\
				\cmidrule(lr){3-5} \cmidrule(lr){6-8}
				&&  Varying  & Constant & Nonzero  & Varying    & Constant & Nonzero      \\
				\textbf{Example 1} &TP	&2.98(0.15)	&4.73(0.45)	&5.00(0.00)	&3.00(0.00)	&0.00(0.00)	&5.00(0.00)		
				\\
				&FP	&1.89(1.40)	&0.42(0.69)	&4.07(2.27)	&6.13(1.18)	&0.00(0.00)	&3.16(2.02)	 			 		
				\\
				\textbf{Example 2}  &TP	&3.00(0.00)	&4.76(0.48)	&5.00(0.00)	&3.00(0.00)	&0.00(0.00)	&5.00(0.00)
				\\
				&FP	&3.27(2.38)	&0.36(0.57)	&5.13(2.32)	&6.78(1.52)	&0.00(0.00)	&4.20(2.21)	
				\\
				\textbf{Example 3} &TP &3.00(0.00)	&4.78(0.42)	&5.00(0.00)	&3.00(0.00)	&0.00(0.00)	&5.00(0.00)
				\\
				&FP	&2.09(1.86)	&0.24(0.53)	&4.33(2.32)	&6.04(1.30)	&0.00(0.00)	&3.42(2.11)
				\\
				\textbf{Example 4} &TP	&3.00(0.00)	&4.78(0.52)	&5.00(0.00)	&3.00(0.00)	&0.00(0.00)	&5.00(0.00)
				\\
				&FP	&3.33(1.98)	&0.24(0.43)	&6.47(2.66)	&6.51(1.36)	&0.00(0.00)	&5.07(2.61)
				\\
				\hline
		\end{tabu} }
	\end{center}
	\centering
\end{table}

\clearpage
\subsection{Estimation and prediction results}

\begin{table} [h!]
	\def\arraystretch{1.2}
	\begin{center}
		\caption[Simulation results in Example 2.]{Simulation results in Example 2. SNP genotype data $(n,p,q)$ = (500, 100, 2). mean(sd) of the integrated mean squared error (IMSE), mean squared error (MSE), total squared errors for all estimates and prediction errors based on 100 replicates.}\label{est2}
		\centering
		\fontsize{10}{12}\selectfont{
			\begin{tabu} to \textwidth{ X[c] X[c] X[c] X[c]  X[c] X[c] }
				\hline
				&  BSSVC-SI  & BSSVC  & BVC-SI & BVC & BL \\
				\hline
				\textbf{IMSE} & & & &   &   \\
				$\beta_{0}(Z)$  &0.043(0.013)	&0.043(0.012)	&0.055(0.024)	&0.055(0.022) &0.810(0.038)										
				\\
				$\beta_{1}(Z)$ &0.042(0.020)	&0.021(0.012)	&0.069(0.031)	&0.085(0.031)	&0.127(0.022)	
				\\
				$\beta_{2}(Z)$ &0.027(0.018)	&0.021(0.012)	&0.044(0.025)	&0.049(0.025)	&0.234(0.037)				
				\\
				$\beta_{3}(Z)$  &0.030(0.026)	&0.026(0.022)	&0.074(0.034)	&0.094(0.038)	&0.256(0.055)
				\\[0.1cm]
				\textbf{MSE} & & & &   &   \\
				$\alpha_{1}$  &0.011(0.012)	&0.012(0.013)	&0.022(0.023)	&0.022(0.022)	&0.010(0.014)		
				\\
				$\alpha_{2}$  &0.003(0.003)	&0.003(0.004)	&0.007(0.009)	&0.007(0.008)	&0.010(0.014)			
				\\
				$\zeta_{0}$ &0.033(0.025)	&0.024(0.019)	&0.081(0.057)	&0.106(0.062)	&0.024(0.032)	
				\\[0.1cm]
				$\zeta_{1}$ &0.005(0.005)	&0.006(0.007)	&0.009(0.013)	&0.008(0.013)	&0.015(0.021)			
				\\
				$\zeta_{2}$  &0.008(0.009)	&0.006(0.008)	&0.019(0.023)	&0.019(0.022)	&0.027(0.037)				
				\\
				$\zeta_{3}$ &0.009(0.015)	&0.009(0.013)	&0.017(0.023)	&0.019(0.025)	&0.053(0.068)				
				\\
				$\zeta_{4}$ &0.009(0.014)	&0.011(0.019)	&0.011(0.015)	&0.010(0.014)	&0.025(0.033)					
				\\
				$\zeta_{5}$  &0.006(0.007)	&0.006(0.008)	&0.020(0.028)	&0.024(0.030)	&0.028(0.030)					
				\\[0.1cm]
				\textbf{Total}  &0.227(0.083)	&0.253(0.104)	&2.020(0.260)	&1.931(0.228)	&4.329(0.436)			
				\\[0.1cm]
				\textbf{Pred.} & & & &   &   \\
				\textbf{Error} &1.160(0.071)	&1.169(0.064)	&2.196(0.180)	&2.155(0.154)	&9.593(0.863)		
				\\[0.1cm]
				\hline
		\end{tabu} }
	\end{center}
	\centering
\end{table}

\begin{table} [h!]
	\def\arraystretch{1.2}
	\begin{center}
		\caption[Simulation results in Example 3.]{Simulation results in Example 3. SNP genotype data based on the linkage disequilibrium (LD) structure $(n,p,q)$ = (500, 100, 2). mean(sd) of the integrated mean squared error (IMSE), mean squared error (MSE), total squared errors for all estimates and prediction errors based on 100 replicates.}\label{est3}
		\centering
		\fontsize{10}{12}\selectfont{
			\begin{tabu} to \textwidth{ X[c] X[c] X[c] X[c]  X[c] X[c] }
				\hline
				&  BSSVC-SI  & BSSVC  & BVC-SI & BVC & BL \\
				\hline
				\textbf{IMSE} & & & &   &   \\
				$\beta_{0}(Z)$  &0.046(0.015)	&0.045(0.014)	&0.060(0.023)	&0.058(0.022)	&0.818(0.038)				
				\\
				$\beta_{1}(Z)$ &0.059(0.025)	&0.025(0.013)	&0.112(0.045)	&0.120(0.045)	&0.136(0.035)		
				\\
				$\beta_{2}(Z)$ &0.035(0.018)	&0.023(0.017)	&0.051(0.024)	&0.054(0.025)	&0.248(0.050)
				\\
				$\beta_{3}(Z)$  &0.032(0.019)	&0.027(0.018)	&0.083(0.049)	&0.105(0.051)	&0.260(0.049)			
				\\[0.1cm]
				\textbf{MSE} & & & &   &   \\
				$\alpha_{1}$  &0.003(0.005)	&0.003(0.004)	&0.006(0.009)	&0.006(0.009)	&0.011(0.015)			
				\\
				$\alpha_{2}$  &0.005(0.006)	&0.004(0.006)	&0.010(0.015)	&0.009(0.013)	&0.011(0.015)			
				\\
				$\zeta_{0}$ &0.010(0.013)	&0.008(0.011)	&0.024(0.034)	&0.023(0.033)	&0.039(0.056)					
				\\[0.1cm]
				$\zeta_{1}$ &0.008(0.014)	&0.008(0.015)	&0.012(0.016)	&0.011(0.014)	&0.022(0.026)			
				\\
				$\zeta_{2}$  &0.010(0.014)	&0.008(0.012)	&0.024(0.035)	&0.025(0.035)	&0.044(0.058)		
				\\
				$\zeta_{3}$ &0.009(0.008)	&0.010(0.009)	&0.024(0.034)	&0.024(0.032)	&0.049(0.052)					
				\\
				$\zeta_{4}$ &0.013(0.017)	&0.022(0.022)	&0.026(0.034)	&0.023(0.030)	&0.044(0.046)			
				\\
				$\zeta_{5}$  &0.017(0.026)	&0.038(0.034)	&0.032(0.039)	&0.030(0.036)	&0.056(0.067)		
				\\[0.1cm]
				\textbf{Total}  &0.307(0.107)	&0.407(0.141)	&2.176(0.219)	&2.015(0.207)	&4.628(0.510)			
				\\[0.1cm]
				\textbf{Pred.} & & & &   &   \\
				\textbf{Error} &1.203(0.064)	&1.209(0.068)	&2.164(0.137)	&2.088(0.132)	&9.483(0.995)
				\\[0.1cm]
				\hline
		\end{tabu} }
	\end{center}
	\centering
\end{table}

\begin{table} [h!]
	\def\arraystretch{1.2}
	\begin{center}
		\caption[Simulation results in Example 4.]{Simulation results in Example 4. SNP genotype from T2D data $(n,p,q)$ = (500, 100, 2). mean(sd) of the integrated mean squared error (IMSE), mean squared error (MSE), total squared errors for all estimates and prediction errors based on 100 replicates.}\label{est4}
		\centering
		\fontsize{10}{12}\selectfont{
			\begin{tabu} to \textwidth{ X[c] X[c] X[c] X[c]  X[c] X[c] }
				\hline
				&  BSSVC-SI  & BSSVC  & BVC-SI & BVC & BL \\
				\hline
				\textbf{IMSE} & & & &   &   \\
				$\beta_{0}$  &0.051(0.019)	&0.051(0.019)	&0.066(0.021)	&0.064(0.020)	&0.809(0.050)			
				\\
				$\beta_{1}(Z)$ &0.032(0.015)	&0.018(0.011)	&0.052(0.027)	&0.068(0.030)	&0.136(0.032)			
				\\
				$\beta_{2}(Z)$&0.015(0.010)	&0.014(0.009)	&0.029(0.021)	&0.033(0.020)	&0.225(0.026)	
				\\
				$\beta_{3}(Z)$ &0.023(0.018)	&0.019(0.013)	&0.051(0.027)	&0.066(0.030)	&0.238(0.039)	
				\\[0.1cm]
				\textbf{MSE} & & & &   &   \\
				$\alpha_{1}$  &0.003(0.003)	&0.003(0.004)	&0.007(0.013)	&0.007(0.013)	&0.010(0.013)		
				\\
				$\alpha_{2}$ &0.003(0.004)	&0.003(0.005)	&0.005(0.005)	&0.005(0.004)	&0.010(0.011)		
				\\
				$\zeta_{0}$ &0.007(0.014)	&0.007(0.016)	&0.015(0.015)	&0.014(0.014)	&0.054(0.086)				
				\\[0.1cm]
				$\zeta_{1}$ &0.004(0.006)	&0.004(0.005)	&0.008(0.013)	&0.008(0.011)	&0.019(0.027)		
				\\
				$\zeta_{2}$  &0.009(0.009)	&0.006(0.007)	&0.019(0.022)	&0.018(0.020)	&0.023(0.031)			
				\\
				$\zeta_{3}$ &0.007(0.010)	&0.007(0.009)	&0.012(0.018)	&0.012(0.021)	&0.019(0.024)		
				\\
				$\zeta_{4}$ &0.004(0.004)	&0.004(0.004)	&0.006(0.008)	&0.006(0.008)	&0.019(0.022)				
				\\
				$\zeta_{5}$  &0.003(0.003)	&0.003(0.004)	&0.013(0.014)	&0.014(0.014)	&0.018(0.023)		
				\\[0.1cm]
				\textbf{Total} &0.178(0.052)	&0.194(0.049)	&1.751(0.194)	&1.648(0.157)  &4.062(0.401)	
				\\[0.1cm]
				\textbf{Pred.} & & & &   &   \\
				\textbf{Error} &1.141(0.073)	&1.147(0.064)	&2.164(0.134)	&2.109(0.125)	&11.337(0.991)
				\\[0.1cm]
				\hline
		\end{tabu} }
	\end{center}
	\centering
\end{table}

\clearpage
\subsection{Sensitivity analysis on smoothness specification}
Let $O$ denotes the degree of B spline basis, and $K$ denotes the number of interior knots. Huang et al. \cite{KO1} and Huang et al. \cite{KO2} show that $n^{1/(2O+3)}$ is the optimal order of the number of spline knots $K$. For quadratic and cubic splines corresponding to $O=2$ and $3$ respectively, we conduct a sensitivity analysis for the proposed model under the setting of Example 2, for $K\in[1,5]$. Table \ref{sens2} shows that $K=1$ leads to unsatisfactory performance, especially for prediction. When $K\geq2$, different values of $K$ lead to similar performance under $O=2$ and $O=3$. This suggests the model performance is insensitive with respect to the smoothness specification. 

\begin{table} [h!]
	\def\arraystretch{1.5}
	\begin{center}
		\caption[Sensitivity analysis on smoothness specification.]{Sensitivity analysis on smoothness specification. $(n,p,q)$ = (500, 100, 2). mean(sd) of true positives (TP), false positives (FP) and prediction error based on 100 replicates.}\label{sens2}
		\centering
		\fontsize{10}{12}\selectfont{
			\begin{tabu} to \textwidth{ X[0.6c] X[c] X[c] X[c] X[c] X[c] X[c] X[c]}
				\hline
				$O=2$ &  \multicolumn{2}{c}{Varying}  & \multicolumn{2}{c}{Constant}  & \multicolumn{2}{c}{Nonzero} & \\
				\hline
				$K$ & TP&  FP& TP&  FP& TP&  FP&  Pred.Error      \\
				\hline
				1 &2.97(0.18)	&0.20(0.55)	&4.87(0.35)	&0.10(0.31)	&4.97(0.18)	&0.03(0.18)	&1.998(0.152)			
				\\
				2 &3.00(0.00)	&0.03(0.18)	&5.00(0.00)	&0.00(0.00)	&5.00(0.00)	&0.00(0.00) &1.172(0.071)		
				\\
				3 &3.00(0.00)	&0.00(0.00)	&5.00(0.00)	&0.03(0.18)	&5.00(0.00)	&0.13(0.43) &1.140(0.093)	
				\\
				4 &3.00(0.00)	&0.07(0.25)	&4.93(0.25)	&0.00(0.00)	&5.00(0.00)	&0.00(0.00) &1.200(0.118)				
				\\
				5 &2.98(0.15)	&0.00(0.00)	&5.00(0.00)	&0.16(0.37)	&5.00(0.00)	&0.07(0.25) &1.150(0.077)
				\\
				\hline
				$O=3$ &  \multicolumn{2}{c}{Varying}  & \multicolumn{2}{c}{Constant}  & \multicolumn{2}{c}{Nonzero} & \\
				\hline
				$K$ & TP&  FP& TP&  FP& TP&  FP&  Pred.Error      \\
				\hline
				1 &3.00(0.00)	&0.17(0.38)	&4.87(0.35)	&0.03(0.18)	&5.00(0.00)	&0.00(0.00)	&1.676(0.159)	
				\\
				2 &3.00(0.00)	&0.00(0.00)	&5.00(0.00)	&0.03(0.18)	&5.00(0.00)	&0.03(0.18) &1.089(0.054)
				\\
				3 &3.00(0.00)	&0.00(0.00)	&4.97(0.18)	&0.03(0.18)	&5.00(0.00)	&0.00(0.00) &1.185(0.072)
				\\
				4 &3.00(0.00)	&0.00(0.00)	&5.00(0.00)	&0.03(0.18)	&5.00(0.00)	&0.03(0.18) &1.156(0.078)
				\\
				5 &2.98(0.15)	&0.11(0.32)	&4.96(0.21)	&0.04(0.21)	&5.00(0.00)	&0.00(0.00) &1.258(0.107)
				\\
				\hline
		\end{tabu} }
	\end{center}
	\centering
\end{table}

\clearpage
\subsection{{Computational cost}}
\begin{table} [h!]
	\def\arraystretch{1.5}
	\begin{center}
		\caption[Computational cost.]{{Computational cost analysis for BSSVC-SI under the setting of Example 1. $p$: number of genes. time: CPU time (in seconds) for 10,000 MCMC iterations. The number of regression coefficients to be estimated after basis expansion is approximately $q_{n}p + p$, where $q_{n}$ is the number of basis function. In this study, $q_{n}=5$.}}\label{cost}
		
		\centering
		\fontsize{10}{12}\selectfont{
			\begin{tabu} to 0.8\textwidth{ X[c]  X[c] X[c] X[c] X[c] X[c]}
				\hline
				\multicolumn{2}{c}{$n=500$} &  \multicolumn{2}{c}{$n=1500$}  &  \multicolumn{2}{c}{$n=3000$}\\
				\cmidrule(lr){1-2}  \cmidrule(lr){3-4} \cmidrule(lr){5-6} 
				\# of genes & time & \# of genes & time & \# of genes & time\\
				\textbf{$p=100$} &11.707	&\textbf{$p=300$}	&121.396 &\textbf{$p=600$}	&552.043	
				\\
				\textbf{$p=200$}  &24.878	&\textbf{$p=600$}	&236.571 &\textbf{$p=900$}	&834.645
				\\
				\textbf{$p=300$} &36.372 &\textbf{$p=900$}	&341.366 &\textbf{$p=1200$} &988.939
				\\
				\hline
		\end{tabu} }
	\end{center}
	\centering
\end{table}

\clearpage
\subsection{{The estimated varying coefficient functions}}
\begin{figure}[h!]
	\centering
	\includegraphics[angle=0,origin=c,width=0.8\textwidth]{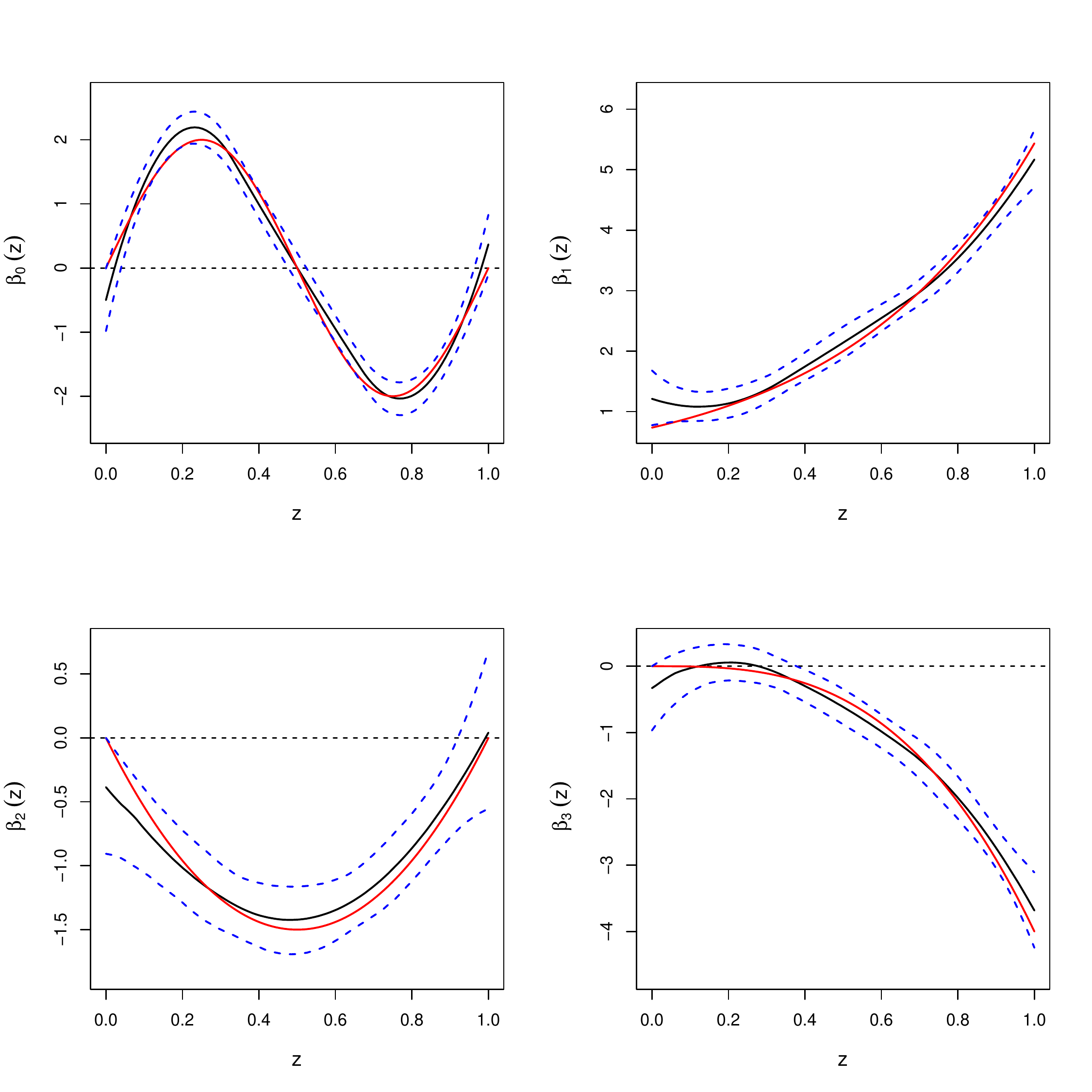}
	\caption[The estimated varying coefficient functions.]{{Simulation study in Example 1 for the proposed method (BSSVC-SI). Red line: true parameter values. Black line: median estimates of varying coefficients for BSSVC-SI. Blue dashed lines: 95\% credible intervals for the estimated varying coefficients.}}
	\label{fig:curve}
\end{figure}

\clearpage
\subsection{{Assessment of the convergence of MCMC chains}}
\begin{figure}[h!]
	\centering
	\includegraphics[angle=0,origin=c,width=1\textwidth]{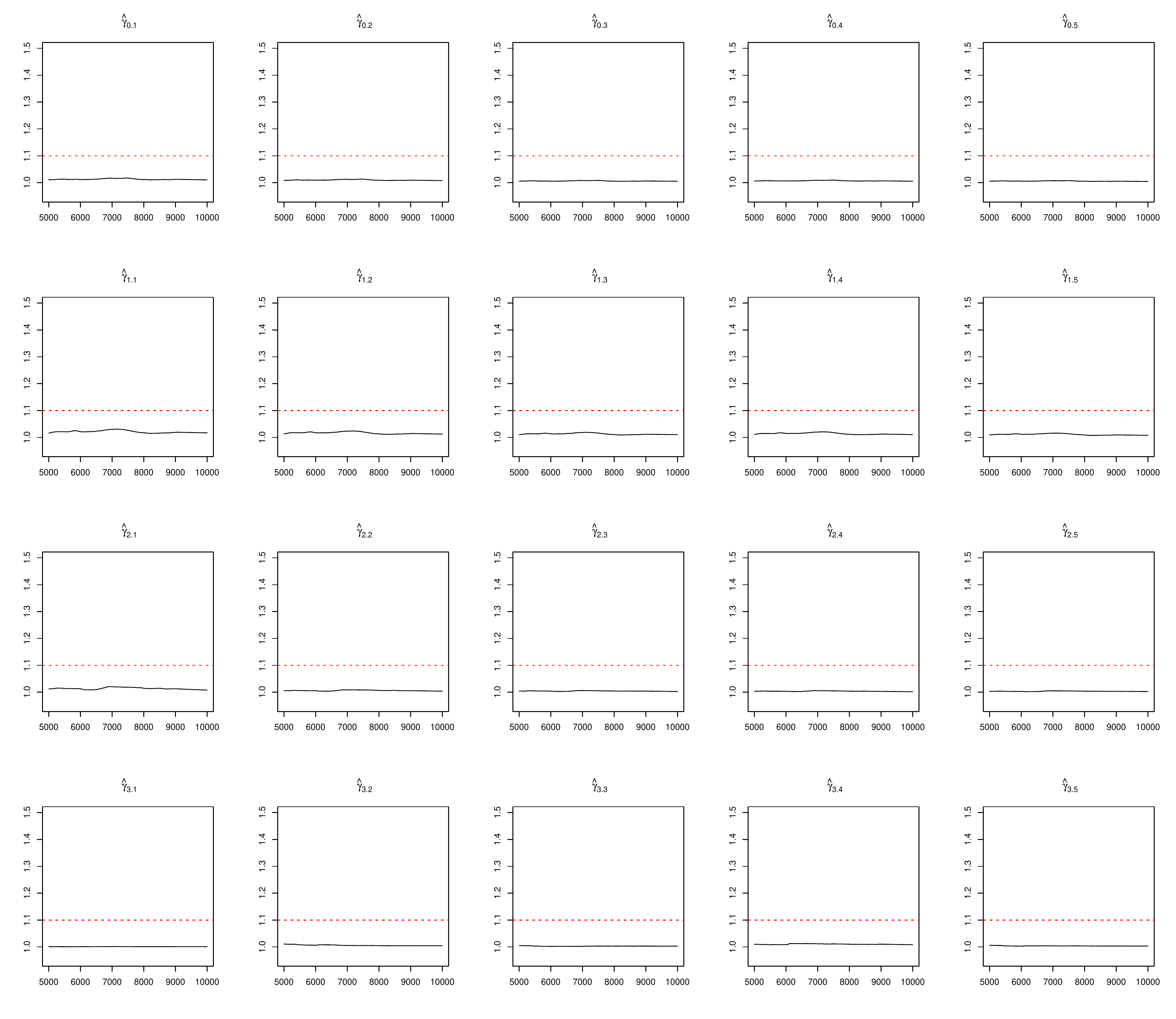}
	\caption[]{{Potential scale reduction factor (PSRF) against iterations for varying coefficient functions in Figure \ref{fig:curve}. Black line: the PSRF. Red line: the threshold of 1.1. The $\hat{\gamma}_{j1}$ to $\hat{\gamma}_{j5}$, $(j=0,\dots,3)$, represent the five estimated spline coefficients for the varying coefficient function $\beta_{j}$, respectively.}}
	\label{fig:cvg}
\end{figure}

\clearpage
\section{Additional results for real data analysis}
\begin{figure}[h!]
	\centering
	\includegraphics[angle=0,origin=c,width=0.9\textwidth]{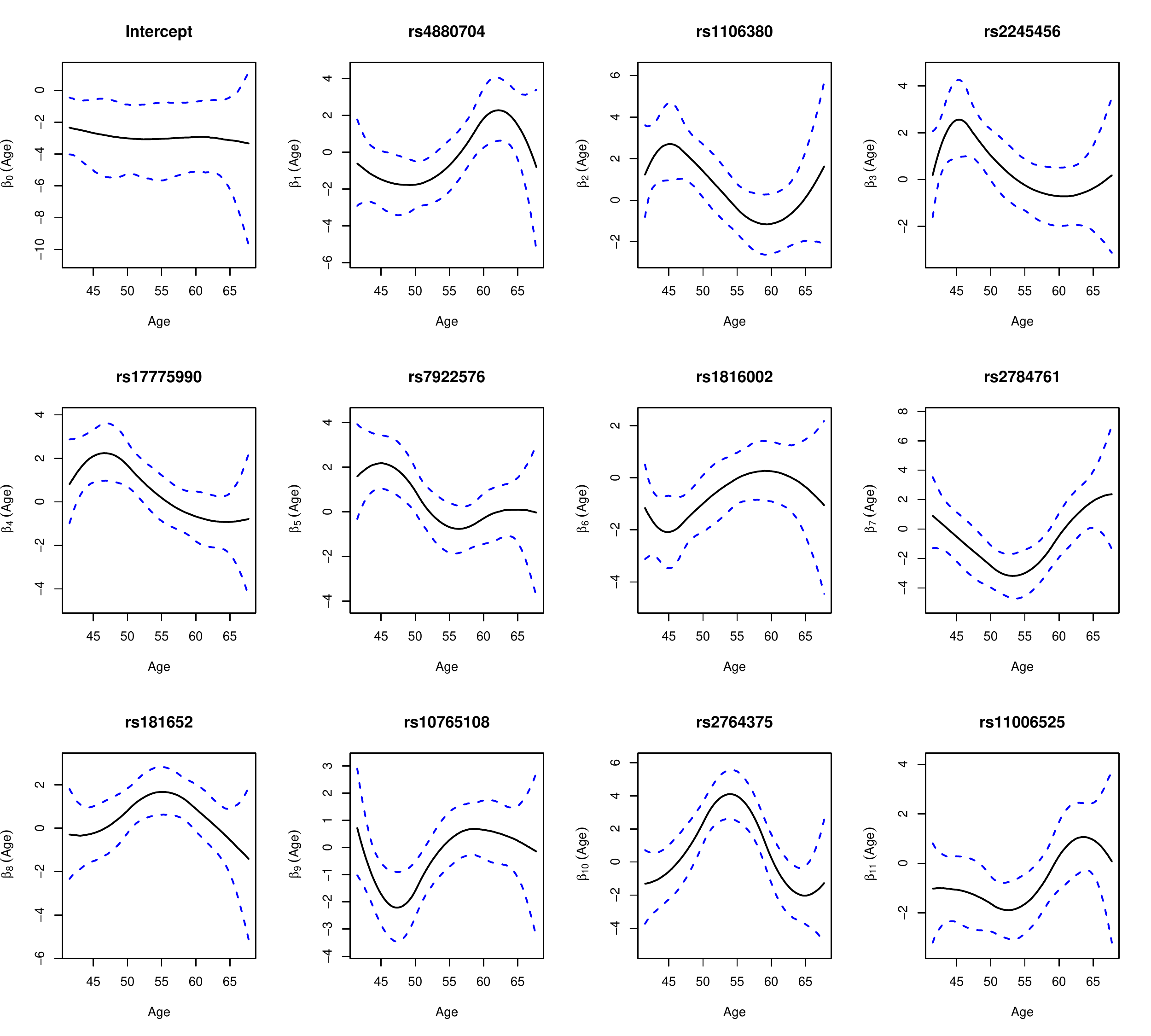}
	\caption[Real data analysis for the proposed method.]{Real data analysis for the proposed method (BSSVC-SI). Black line: median estimates of varying coefficients for BSSVC-SI. Blue dashed lines: 95\% credible intervals for the estimated varying coefficients.}
	\label{fig:curve2}
\end{figure}

\clearpage
\section{Posterior inference}
\subsection{Posterior inference for the BSSVC-SI method}
\label{makereferenceA.7}
\subsubsection{Priors}
\begin{equation*}
\begin{aligned}
Y|\eta, \gamma_{11},\dots,\gamma_{p1},& \gamma_{1*},\dots,\gamma_{p*},\alpha_{1},\dots,\alpha_{q},\zeta_{0}, \zeta_{1},\dots,\zeta_{p},\sigma^{2} \\ 
& \propto (\sigma^{2})^{-\frac{n}{2}} \exp\Big\{-\frac{1}{2\sigma^{2}}(Y-\mu)^\top (Y-\mu)\Big\}
\end{aligned}
\end{equation*}

\begin{equation*}
\eta\thicksim \text{N}_{q_{n}}(0, \, \Sigma_{\eta0})
\end{equation*}
\begin{equation*}
\alpha\thicksim \text{N}_{q}(0, \, \Sigma_{\alpha0})
\end{equation*}
\begin{equation*}
\zeta_{0}\thicksim \text{N}(0, \, \sigma_{\zeta_{0}}^{2})
\end{equation*}

\begin{equation*}
\gamma_{j1}|\pi_{c}, \tau_{cj}^{2}, \sigma^{2} \thicksim \pi_{c} \text{N}(0, \, \sigma^{2}\tau_{cj}^{2}) + (1-\pi_{c})\delta_{0}(\gamma_{j1}), \quad j=1,\dots, p
\end{equation*}
\begin{equation*}
\tau_{cj}^{2}|\lambda_{c} \thicksim \frac{\lambda_{c}^{2}}{2}\exp (-\frac{\lambda_{c}^{2}\tau_{cj}^{2}}{2}), \quad j=1,\dots, p
\end{equation*}

\begin{equation*}
\gamma_{j*}|\pi_{v}, \tau_{vj}^{2}, \sigma^{2} \thicksim \pi_{v} \text{N}_{L}(0, \, \text{Diag}(\sigma^{2}\tau_{vj}^{2},\ldots, \sigma^{2}\tau_{vj}^{2})) + (1-\pi_{v})\delta_{0}(\gamma_{j*}), \quad j=1,\dots, p
\end{equation*}
\begin{equation*}
\tau_{vj}^{2}|\lambda_{v} \thicksim \text{Gamma}(\frac{L+1}{2}, \, \frac{L\lambda_{v}^{2}}{2}), \quad j=1,\dots, p
\end{equation*}

\begin{equation*}
\zeta_{j}|\pi_{e}, \tau_{ej}^{2}, \sigma^{2} \thicksim \pi_{e} \text{N}(0, \, \sigma^{2}\tau_{ej}^{2})+ (1-\pi_{e})\delta_{0}(\zeta_{j}), \quad j=1,\dots, p
\end{equation*}
\begin{equation*}
\tau_{ej}^{2}|\lambda_{e} \thicksim \frac{\lambda_{e}^{2}}{2}\exp (-\frac{\lambda_{e}^{2}\tau_{ej}^{2}}{2}), \quad j=1,\dots, p
\end{equation*}

\begin{equation*}
\sigma^{2} \thicksim (\sigma^{2})^{-s-1}\exp(-\frac{h}{\sigma^{2}})
\end{equation*}

Consider the following conjugate gamma priors for $\lambda_{c}^2$, $\lambda_{v}^{2}$ and $\lambda_{e}^{2}$
\begin{equation*}
\lambda_{c}^{2} \thicksim \text{Gamma}(a_{c}, \, b_{c}), \quad \lambda_{v}^{2} \thicksim \text{Gamma}(a_{v}, \, b_{v}) \quad \text{and}\quad \lambda_{e}^{2} \thicksim \text{Gamma}(a_{e}, \, b_{e})
\end{equation*}

and conjugate beta priors for $\pi_{c}$, $\pi_{v}$ and $\pi_{e}$
\begin{equation*}
\pi_{c} \thicksim \text{Beta}(r_{c}, \, w_{c}), \quad \pi_{v} \thicksim \text{Beta}(r_{v}, \, w_{v}) \quad \text{and}\quad \pi_{e} \thicksim \text{Beta}(r_{e}, \, w_{e})
\end{equation*}

\subsubsection{Gibbs Sampler}

\begin{equation*}
\setlength{\jot}{10pt}
\begin{aligned}
\pi(\eta|& \text{rest}) \\
& \propto \pi(\eta)\pi(y|\cdot) \\
& \propto \exp\Big(-\frac{1}{2}\eta^\top \Sigma_{\eta0}^{-1}\eta\Big)\exp\Big(-\frac{1}{2\sigma^{2}}(Y-\mu)^\top (Y-\mu)\Big) \\
& \propto \exp\Bigg(-\frac{1}{2}\eta^\top \Sigma_{\eta0}^{-1}\eta -\frac{1}{2\sigma^{2}}(Y-B_{0}\eta-\mu_{(-\eta)})^\top (Y-B_{0}\eta-\mu_{(-\eta)})\Bigg) \\
& \propto \exp\Bigg(\eta^\top \Sigma_{\eta0}^{-1}\eta + \frac{1}{\sigma^{2}}\eta^\top B_{0}^\top B_{0}\eta - \frac{2}{\sigma^{2}}(Y-\mu_{(-\eta)})^\top (B_{0}\eta)\Bigg) \\
& \propto \exp\Bigg(\eta^\top \Big(\Sigma_{\eta0}^{-1}+ \frac{1}{\sigma^{2}}B_{0}^\top B_{0}\Big)\eta -\frac{2}{\sigma^{2}}(Y-\mu_{(-\eta)})^\top B_{0}\eta\Bigg)
\end{aligned}
\end{equation*}
where $B_{0} = (B_{0}(Z_{1}), \dots, B_{0}(Z_{n}))^\top $. Hence, the full conditional distribution of $\eta$ is multivariate normal with mean
\begin{equation*}
\mu_{\eta}=\Big(\Sigma_{\eta0}^{-1}+ \frac{1}{\sigma^{2}}B_{0}^\top B_{0}\Big)^{-1} \Big(\frac{1}{\sigma^{2}}(Y-\mu_{(-\eta)})^\top B_{0}\Big)^\top 
\end{equation*}
and variance
\begin{equation*}
\Sigma_{\eta}=\Big(\Sigma_{\eta0}^{-1}+ \frac{1}{\sigma^{2}}B_{0}^\top B_{0}\Big)^{-1}
\end{equation*}
Similarly, the full conditional distribution of $\alpha$ is $N(\mu_{\alpha}, \Sigma_{\alpha})$ with
\begin{equation*}
\mu_{\alpha}=\Big(\Sigma_{\alpha0}^{-1}+ \frac{1}{\sigma^{2}}W^\top W\Big)^{-1} \Big(\frac{1}{\sigma^{2}}(Y-\mu_{(-\alpha)})^\top W\Big)^\top 
\end{equation*}
and variance
\begin{equation*}
\Sigma_{\alpha}=\Big(\Sigma_{\alpha0}^{-1}+ \frac{1}{\sigma^{2}}W^\top W\Big)^{-1}
\end{equation*}
where $W = (W_{1}, \dots, W_{n})^\top $. And the full conditional distribution of $\zeta_{0}$ is $N(\mu_{\zeta_{0}}, \Sigma_{\zeta_{0}})$ with
\begin{equation*}
\mu_{\zeta_{0}}=\Big(1/\sigma_{\zeta_{0}}^{2}+\frac{1}{\sigma^{2}}\sum_{i=1}^{n}E_{i}^{2}\Big)^{-1} \Big(\frac{1}{\sigma^{2}}\sum_{i=1}^{n}(y_{i}-\mu_{(-\zeta_{0})})E_{i}\Big)
\end{equation*}
and variance
\begin{equation*}
\Sigma_{\zeta_{0}}=\Big(1/\sigma_{\zeta_{0}}^{2} + \frac{1}{\sigma^{2}}\sum_{i=1}^{n}E_{i}^{2}\Big)^{-1}
\end{equation*}
The full conditional distribution of $\gamma_{j*}$ 
\begin{equation}\label{equr:rstar}
\setlength{\jot}{10pt}
\begin{aligned}
\pi(\gamma_{j*}|& \text{rest}) \\
& \propto \pi(\gamma_{j*}|\tau_{vj}^{2}, \sigma^{2})\pi(y|\cdot) \\
& \propto (\sigma^{2})^{-\frac{n}{2}} \exp\Bigg(- \frac{1}{2\sigma^{2}} (Y-U_{j}\gamma_{j*}-\mu_{(-\gamma_{j*})})^\top (Y-U_{j}\gamma_{j*}-\mu_{(-\gamma_{j*})}) \Bigg) \\
& \times \Bigg( \pi_{v}(2\pi\sigma^{2}\tau_{vj}^{2})^{-\frac{L}{2}} \exp\Big( -\frac{1}{2\sigma^{2}\tau_{vj}^{2}}\gamma_{j*}^\top \gamma_{j*}\Big) \textbf{I}_{\{\gamma_{j*} \neq 0\}}+ (1-\pi_{v})\delta_{0}(\gamma_{j*}) \Bigg)\\
\end{aligned}
\end{equation}
where $U_{j} = (U_{1j}, \dots, U_{nj})^\top $ is a $n\times L$ matrix. 
Let $\Sigma_{\gamma_{j*}} = (U_{j}^\top U_{j} + \frac{1}{\tau_{vj}^{2}}\textbf{I}_{L})^{-1}$, we have
\begin{equation*}
\setlength{\jot}{10pt}
\begin{aligned}
l_{vj} &= \pi(\gamma_{j*} \neq 0 | \text{rest}) \\
&= \frac{\pi_{v}}{\pi_{v} + (1-\pi_{v})(\tau_{vj}^{2})^{\frac{L}{2}}|\Sigma_{\gamma_{j*}}|^{-\frac{1}{2}} \exp\Big(- \frac{1}{2\sigma^{2}} \lVert\Sigma_{\gamma_{j*}}^{\frac{1}{2}}U_{j}^\top(Y-\mu_{(-\gamma_{j*})})\rVert_{2}^{2} \Big)} 
\end{aligned}
\end{equation*}
Hence, the full conditional distribution of $\gamma_{j*}$ is a spike and slab distribution
\begin{equation*}
\gamma_{j*}|\text{rest} \thicksim l_{vj} \text{N}(\mu_{\gamma_{j*}}, \, \sigma^{2}\Sigma_{\gamma_{j*}}) + (1-l_{vj})\delta_{0}(\gamma_{j*})
\end{equation*}
with mean
\begin{equation*}\label{equr:meanrs}
\mu_{\gamma_{j*}}=\Sigma_{\gamma_{j*}}U_{j}^\top (Y - \mu_{(-\gamma_{j*})})
\end{equation*}
This posterior distribution is a mixture of a multivariate normal and a point mass at $0$. To sample from this posterior distribution at the $g$th iteration, we follow the steps:
\begin{itemize}
	\item Generate $u$ from Unif[0,1] 
	\item If $u\leq l_{vj}$
	\begin{itemize}
		\item Generate $t$ from $\text{N}(\mu_{\gamma_{j*}}, \, \sigma^{2}\Sigma_{\gamma_{j*}})$
		\item set $\gamma_{j*}^{(g)}$ = $t$ and $\phi_{vj}^{(g)}$ = $1$
	\end{itemize}
	\item If $u > l_{vj}$
	\begin{itemize}
		\item set $\gamma_{j*}^{(g)}$ = $0$ and $\phi_{vj}^{(g)}$ = $0$
	\end{itemize}
\end{itemize}
Note that, when we sample $\gamma_{j*}^{(g)}$, we also compute the value of $\phi_{vj}^{(g)}$.
The full conditional distribution of $\gamma_{j1}$ can be expressed as
\begin{equation}\label{equr:pr0}
\setlength{\jot}{10pt}
\begin{aligned}
\pi(\gamma_{j1}|& \text{rest}) \\
& \propto \pi(\gamma_{j1}|\tau_{cj}^{2}, \sigma^{2})\pi(y|\cdot) \\
& \propto (\sigma^{2})^{-\frac{n}{2}} \exp\Bigg(- \frac{1}{2\sigma^{2}} (Y-X_{j}\gamma_{j1}-\mu_{(-\gamma_{j1})})^\top (Y-X_{j}\gamma_{j1}-\mu_{(-\gamma_{j1})}) \Bigg) \\
& \times \Bigg( \pi_{c}(2\pi\sigma^{2}\tau_{cj}^{2})^{-\frac{1}{2}} \exp\Big( -\frac{1}{2\sigma^{2}\tau_{cj}^{2}}\gamma_{j1}^{2}\Big) \textbf{I}_{\{\gamma_{j1} \neq 0\}}+ (1-\pi_{c})\delta_{0}(\gamma_{j1}) \Bigg)
\end{aligned}
\end{equation}
Let $\Sigma_{\gamma_{j1}} = (X_{j}^\top X_{j} + \frac{1}{\tau_{cj}^{2}})^{-1}$, we have
\begin{equation*}
\setlength{\jot}{10pt}
\begin{aligned}
l_{cj} &= \pi(\gamma_{j1} \neq 0 | \text{rest}) \\
&= \frac{\pi_{c}}{\pi_{c} + (1-\pi_{c})(\tau_{cj}^{2})^{\frac{1}{2}}(\Sigma_{\gamma_{j1}})^{-\frac{1}{2}} \exp\Big(- \frac{\Sigma_{\gamma_{j1}}}{2\sigma^{2}} \lVert(Y-\mu_{(-\gamma_{j1})})^\top X_{j}\rVert_{2}^{2} \Big)} 
\end{aligned}
\end{equation*}
Hence, the full conditional distribution of $\gamma_{j1}$ is a spike and slab distribution
\begin{equation*}
\gamma_{j1}|\text{rest} \thicksim l_{cj} N(\mu_{\gamma_{j1}}, \, \sigma^{2}\Sigma_{\gamma_{j1}}) + (1-l_{cj})\delta_{0}(\gamma_{j1})
\end{equation*}
with mean
\begin{equation*}\label{equr:meanr0}
\mu_{\gamma_{j1}}=\Sigma_{\gamma_{j1}}  X_{j}^\top (Y - \mu_{(-\gamma_{j1})})
\end{equation*}
The full conditional distribution of $\zeta_{j}, j=1,\dots,p$
\begin{equation}\label{equr:pzeta}
\setlength{\jot}{10pt}
\begin{aligned}
\pi(\zeta_{j}|& \text{rest}) \\
& \propto \pi(\zeta_{j}|\tau_{j}^{2}, \sigma^{2})\pi(y|\cdot) \\
& \propto  (\sigma^{2})^{-\frac{n}{2}} \exp\Big(- \frac{1}{2\sigma^{2}} (Y-T_{j}\zeta_{j}-\mu_{(-\zeta_{j})})^\top (Y-T_{j}\zeta_{j}-\mu_{(-\zeta_{j})}) \Big) \\
& \times \Bigg( \pi_{e}(2\pi\sigma^{2}\tau_{ej}^{2})^{-\frac{1}{2}} \exp\Big( -\frac{1}{2\sigma^{2}\tau_{ej}^{2}}\zeta_{j}^{2}\Big) \textbf{I}_{\{\zeta_{j} \neq 0\}}+ (1-\pi_{e})\delta_{0}(\zeta_{j}) \Bigg)
\end{aligned}
\end{equation}
Let $\Sigma_{\zeta_{j}} = (T_{j}^\top T_{j} + \frac{1}{\tau_{ej}^{2}})^{-1}$, we have
\begin{equation*}
\setlength{\jot}{10pt}
\begin{aligned}
l_{ej} &= \pi(\zeta_{j} \neq 0 | \text{rest}) \\
&= \frac{\pi_{e}}{\pi_{e} + (1-\pi_{e})(\tau_{ej}^{2})^{\frac{1}{2}}(\Sigma_{\zeta_{j}})^{-\frac{1}{2}} \exp\Big(- \frac{\Sigma_{\zeta_{j}}}{2\sigma^{2}} \lVert(Y-\mu_{(-\zeta_{j})})^\top T_{j}\rVert_{2}^{2} \Big)} 
\end{aligned}
\end{equation*}
Hence, the full conditional distribution of $\zeta_{j}$ is a spike and slab distribution
\begin{equation*}
\zeta_{j}|\text{rest} \thicksim l_{ej} N(\mu_{\zeta_{j}}, \, \sigma^{2}\Sigma_{\zeta_{j}}) + (1-l_{ej})\delta_{0}(\zeta_{j})
\end{equation*}
where
\begin{equation*}
\mu_{\zeta_{j}}=\Sigma_{\zeta_{j}} T_{j}^\top(Y-\mu_{(-\zeta_{j})}) 
\end{equation*}
Now, we derive the full conditional distribution for $\tau_{vj}^{2}$, $\tau_{cj}^{2}$ and $\tau_{ej}^{2}$.
\begin{equation}\label{equr:ptaus}
\setlength{\jot}{10pt}
\begin{aligned}
\pi(\tau_{vj}^{2}|& \text{rest}) \\
& \propto \pi(\tau_{vj}^{2}|\lambda_{v}) \pi(\gamma_{j*}|\tau_{vj}^{2}, \sigma^{2}) \\
& \propto (\tau_{vj}^{2})^{\frac{L+1}{2}-1}\exp\Big(-\tau_{vj}^{2}\frac{L\lambda_{v}^{2}}{2}\Big)  \\
&\times \Bigg( \pi_{v}(2\pi\sigma^{2}\tau_{vj}^{2})^{-\frac{L}{2}} \exp\Big( -\frac{1}{2\sigma^{2}\tau_{vj}^{2}}\gamma_{j*}^\top \gamma_{j*}\Big) \textbf{I}_{\{\gamma_{j*} \neq 0\}}+ (1-\pi_{v})\delta_{0}(\gamma_{j*}) \Bigg)
\end{aligned}
\end{equation}
When $\gamma_{j*} = 0$, \ref{equr:ptaus} is equal to 
\begin{equation*}
\setlength{\jot}{10pt}
(1-\pi_{v}) (\tau_{vj}^{2})^{\frac{L+1}{2}-1}\exp\Big(-\tau_{vj}^{2}\frac{L\lambda_{v}^{2}}{2}\Big) 
\end{equation*}
Therefore, when $\gamma_{j*} = 0$, the posterior distribution for $(\tau_{vj}^{2})^{-1}$ is Inverse-Gamma($\frac{L+1}{2}$,\, $\frac{L\lambda_{v}^{2}}{2}$). When $\gamma_{j*} \neq 0$, \ref{equr:ptaus} is equal to 
\begin{equation*}\label{equr:ptaus1}
\setlength{\jot}{10pt}
\begin{aligned}
\pi(\tau_{vj}^{2}|& \text{rest}) \\
& \propto (1-\pi_{v})(2\pi\sigma^{2}\tau_{vj}^{2})^{-\frac{L}{2}}  (\tau_{vj}^{2})^{\frac{L+1}{2}-1}\exp\Big(-\tau_{vj}^{2}\frac{L\lambda_{v}^{2}}{2}\Big) \exp\Big( -\frac{1}{2\sigma^{2}\tau_{vj}^{2}}\gamma_{j*}^\top \gamma_{j*}\Big) \\
& \propto (1-\pi_{v})(2\pi\sigma^{2})^{-\frac{L}{2}} (\tau_{vj}^{2})^{-\frac{1}{2}} \exp\Bigg(-\tau_{vj}^{2}\frac{L\lambda_{v}^{2}}{2} -\frac{\lVert\gamma_{j*}\rVert_{2}^{2}}{2\sigma^{2}\tau_{vj}^{2}}\Bigg) \\
\end{aligned}
\end{equation*}
Therefore, when $\gamma_{j*}\neq 0$, the posterior distribution for $(\tau_{vj}^{2})^{-1}$ is Inverse-Gaussian($L\lambda_{v}^{2}$, $\sqrt{\frac{L\lambda_{v}^{2}\sigma^{2}}{\lVert\gamma_{j*}\rVert_{2}^{2}}}$). Together 
\begin{equation*}
(\tau_{vj}^{2})^{-1}|\text{rest} \thicksim \begin{cases}
\scalebox{1}{Inverse-Gamma($\frac{L+1}{2}$,\, $\frac{L\lambda_{v}^{2}}{2}$)}& { \text{if} \; \gamma_{j*} = 0} \\[6pt]
\scalebox{1}{Inverse-Gaussian($L\lambda_{v}^{2}$, $\sqrt{\frac{L\lambda_{v}^{2}\sigma^{2}}{\lVert\gamma_{j*}\rVert_{2}^{2}}}$)}& { \text{if} \; \gamma_{j*} \neq 0}
\end{cases}
\end{equation*}
Similarly, the posterior distribution for $(\tau_{cj}^{2})^{-1}$ is
\begin{equation}\label{equr:ptau00}
\setlength{\jot}{10pt}
\begin{aligned}
\pi(\tau_{cj}^{2}|& \text{rest}) \\
& \propto \pi(\tau_{cj}^{2}|\lambda_{c}) \pi(\gamma_{j1}|\tau_{cj}^{2}, \sigma^{2}) \\
& \propto \frac{\lambda_{c}^{2}}{2} \exp\Big(-\tau_{cj}^{2}\frac{\lambda_{c}^{2}}{2}\Big)  \\
& \times \Bigg( \pi_{c}(2\pi\sigma^{2}\tau_{cj}^{2})^{-\frac{1}{2}} \exp\Big( -\frac{1}{2\sigma^{2}\tau_{cj}^{2}}\gamma_{j1}^{2}\Big) \textbf{I}_{\{\gamma_{j1} \neq 0\}}+ (1-\pi_{c})\delta_{0}(\gamma_{j1}) \Bigg)
\end{aligned}
\end{equation}
When $\gamma_{j1} = 0$, \ref{equr:ptau00} is equal to 
\begin{equation*}
\setlength{\jot}{10pt}
(1-\pi_{c}) \frac{\lambda_{c}^{2}}{2} \exp\Big(-\tau_{cj}^{2}\frac{\lambda_{c}^{2}}{2}\Big) 
\end{equation*}
Therefore, when $\gamma_{j1} = 0$, the posterior distribution for $(\tau_{cj}^{2})^{-1}$ is Inverse-Gamma($1$,\, $\frac{\lambda_{c}^{2}}{2}$). When $\gamma_{j1} \neq 0$, \ref{equr:ptau00} is equal to 
\begin{equation*}
\setlength{\jot}{10pt}
\begin{aligned}
\pi(\tau_{cj}^{2}|& \text{rest}) \\
& \propto (1-\pi_{c})(2\pi\sigma^{2}\tau_{cj}^{2})^{-\frac{1}{2}}  {\frac{\lambda_{c}^{2}}{2}}\exp\Big(-\tau_{cj}^{2}\frac{\lambda_{c}^{2}}{2}\Big) \exp\Big( -\frac{1}{2\sigma^{2}\tau_{cj}^{2}}\gamma_{j1}^{2}\Big) \\
& \propto (\tau_{cj}^{2})^{-\frac{1}{2}} \exp\Bigg(-\tau_{cj}^{2}\frac{\lambda_{c}^{2}}{2} -\frac{\gamma_{j1}^{2}}{2\sigma^{2}\tau_{cj}^{2}}\Bigg) \\
\end{aligned}
\end{equation*}
Therefore, when $\gamma_{j1}\neq 0$, the posterior distribution for $(\tau_{cj}^{2})^{-1}$ is Inverse-Gaussian($\lambda_{c}^{2}$, $\sqrt{\frac{\lambda_{c}^{2}\sigma^{2}}{\gamma_{j1}^{2}}}$). Together 
\begin{equation*}
(\tau_{cj}^{2})^{-1}|\text{rest} \thicksim \begin{cases}
\scalebox{1}{Inverse-Gamma($1$,\, $\frac{\lambda_{c}^{2}}{2}$)}& { \text{if} \; \gamma_{j1} = 0} \\[6pt]
\scalebox{1}{Inverse-Gaussian($\lambda_{c}^{2}$, $\sqrt{\frac{\lambda_{c}^{2}\sigma^{2}}{\gamma_{j1}^{2}}}$)}& { \text{if} \; \gamma_{j1} \neq 0}
\end{cases}
\end{equation*}
The posterior distribution $(\tau_{ej}^{2})^{-1}$
\begin{equation*}
\setlength{\jot}{10pt}
\begin{aligned}
\pi(\tau_{ej}^{2}|& \text{rest}) \\
& \propto \pi(\tau_{ej}^{2}|\lambda_{e}) \pi(\zeta_{j}|\tau_{ej}^{2}, \sigma^{2}) \\
& \propto \frac{\lambda_{e}^{2}}{2} \exp\Big(-\tau_{ej}^{2}\frac{\lambda_{e}^{2}}{2}\Big)  \\
& \times \Bigg( \pi_{e}(2\pi\sigma^{2}\tau_{ej}^{2})^{-\frac{1}{2}} \exp\Big( -\frac{1}{2\sigma^{2}\tau_{ej}^{2}}\zeta_{j}^{2}\Big) \textbf{I}_{\{\zeta_{j} \neq 0\}}+ (1-\pi_{e})\delta_{0}(\zeta_{j}) \Bigg)
\end{aligned}
\end{equation*}
Following the similar arguments, we have
\begin{equation*}
(\tau_{ej}^{2})^{-1}|\text{rest} \thicksim \begin{cases}
\scalebox{1}{Inverse-Gamma($1$,\, $\frac{\lambda_{e}^{2}}{2}$)}& { \text{if} \; \zeta_{j} = 0} \\[6pt]
\scalebox{1}{Inverse-Gaussian($\lambda_{e}^{2}$, $\sqrt{\frac{\lambda_{e}^{2}\sigma^{2}}{\zeta_{j}^{2}}}$)}& { \text{if} \; \zeta_{j} \neq 0}
\end{cases}
\end{equation*}
Now, we derive the full conditional distribution for $\lambda_{v}^{2}$ and $\tau_{cj}^{2}$. The posterior distribution for $\lambda_{v}^{2}$:
\begin{equation*}
\setlength{\jot}{10pt}
\begin{aligned}
\pi(\lambda_{v}^{2}|& \text{rest}) \\
& \propto \pi(\lambda_{v}^{2})\prod_{j=1}^{p}\pi(\tau_{vj}^{2}|\lambda_{v}^{2}) \\
& \propto (\lambda_{v}^{2})^{a_{v}-1}\exp(-b_{v}\lambda_{v}^{2} )\prod_{j=1}^{p}\Bigg(\frac{L\lambda_{v}^{2}}{2}\Bigg)^{\frac{L+1}{2}}\exp\Bigg(-\frac{L\lambda_{v}^{2}}{2}\tau_{vj}^{2}\Bigg)\\
& \propto (\lambda_{v}^{2})^{a_{v}+\frac{p(L+1)}{2}-1} \exp\Bigg(-(b_{v}+\frac{L\sum_{j=1}^{p}\tau_{vj}^{2}}{2})\lambda_{v}^{2}\Bigg)
\end{aligned}
\end{equation*}
the posterior distribution for $\lambda_{v}^{2}$ is Inverse-Gamma($a_{v}+\frac{p(L+1)}{2}$, $b_{v}+\frac{L\sum_{j=1}^{p}\tau_{vj}^{2}}{2}$).
\begin{equation*}
\setlength{\jot}{10pt}
\begin{aligned}
\pi(\lambda_{c}^{2}|& \text{rest}) \\
& \propto \pi(\lambda_{c}^{2})\prod_{j=1}^{p}\pi(\tau_{cj}^{2}|\lambda_{c}^{2}) \\
& \propto (\lambda_{c}^{2})^{a_{c}-1}\exp(-b_{c}\lambda_{c}^{2} )\prod_{j=1}^{p}\frac{\lambda_{c}^{2}}{2}\exp\Bigg(-\frac{\lambda_{c}^{2}}{2}\tau_{cj}^{2}\Bigg)\\
& \propto (\lambda_{c}^{2})^{a_{c}+p-1} \exp\Bigg(-(b_{c}+\frac{\sum_{j=1}^{p}\tau_{cj}^{2}}{2})\lambda_{c}^{2}\Bigg)
\end{aligned}
\end{equation*}
the posterior distribution for $\lambda_{c}^{2}$ is Inverse-Gamma($a_{c}+p$, $b_{c}+\frac{\sum_{j=1}^{p}\tau_{cj}^{2}}{2}$). Similarly, the full conditional distribution for $\lambda_{e}^{2}$ is Inverse-Gamma($a_{e}+p$, $b_{e}+\frac{\sum_{j=1}^{p}\tau_{ej}^{2}}{2}$).

Next, we derive the full conditional distribution for $\pi_{v}$, $\pi_{c}$ and $\pi_{e}$. The posterior distribution for $\pi_{v}$
\begin{equation*}
\setlength{\jot}{10pt}
\begin{aligned}
\pi(\pi_{v}|& \text{rest}) \\
& \propto \pi(\pi_{v})\prod_{j=1}^{p}\pi(\gamma_{j*}^{2}|\pi_{v}, \tau_{vj}^{2}, \sigma^{2}) \\
& \propto \pi_{v}^{r_{v}-1}(1-\pi_{v})^{w_{v}-1} \\
& \times \prod_{j=1}^{p} \Bigg( \pi_{v}(2\pi\sigma^{2}\tau_{vj}^{2})^{-\frac{L}{2}} \exp\Big( -\frac{1}{2\sigma^{2}\tau_{vj}^{2}}\gamma_{j*}^\top \gamma_{j*}\Big) \textbf{I}_{\{\gamma_{j*} \neq 0\}}+ (1-\pi_{v})\delta_{0}(\gamma_{j*}) \Bigg) \\
& \propto \pi_{v}^{r_{v}+\sum_{j=1}^{p}(\delta_{0}(\gamma_{j*}))-1}(1-\pi_{v})^{w_{v}+\sum_{j=1}^{p}\textbf{I}_{\{\gamma_{j*} \neq 0\}}-1}
\end{aligned}
\end{equation*}
the posterior distribution for $\pi_{v}$ is Beta($r_{v}+\sum_{j=1}^{p}(\delta_{0}(\gamma_{j*}))$, $w_{v}+\sum_{j=1}^{p}\textbf{I}_{\{\gamma_{j*} \neq 0\}}$).
\begin{equation*}
\setlength{\jot}{10pt}
\begin{aligned}
\pi(\pi_{c}|& \text{rest}) \\
& \propto \pi(\pi_{c})\prod_{j=1}^{p}\pi(\gamma_{j1}^{2}|\pi_{c}, \tau_{cj}^{2}, \sigma^{2}) \\
& \propto \pi_{c}^{r_{c}-1}(1-\pi_{c})^{w_{c}-1} \\
& \times \prod_{j=1}^{p} \Bigg( \pi_{c}(2\pi\sigma^{2}\tau_{cj}^{2})^{-\frac{1}{2}} \exp\Big( -\frac{1}{2\sigma^{2}\tau_{cj}^{2}}\gamma_{j1}^{2}\Big) \textbf{I}_{\{\gamma_{j1} \neq 0\}}+ (1-\pi_{c})\delta_{0}(\gamma_{j1}) \Bigg) \\
& \propto \pi_{c}^{r_{c}+\sum_{j=1}^{p}(\delta_{0}(\gamma_{j1}))-1}(1-\pi_{c})^{w_{c}+\sum_{j=1}^{p}\textbf{I}_{\{\gamma_{j1} \neq 0\}}-1}
\end{aligned}
\end{equation*}
the posterior distribution for $\pi_{c}$ is Beta($r_{c}+\sum_{j=1}^{p}(\delta_{0}(\gamma_{j1}))$, $w_{c}+\sum_{j=1}^{p}\textbf{I}_{\{\gamma_{j1} \neq 0\}}$).
\begin{equation*}
\setlength{\jot}{10pt}
\begin{aligned}
\pi(\pi_{e}|& \text{rest}) \\
& \propto \pi(\pi_{e})\prod_{j=1}^{p}\pi(\zeta_{j}^{2}|\pi_{e}, \tau_{ej}^{2}, \sigma^{2}) \\
& \propto \pi_{e}^{r_{e}-1}(1-\pi_{e})^{w_{e}-1} \\
& \times \prod_{j=1}^{p} \Bigg( \pi_{e}(2\pi\sigma^{2}\tau_{ej}^{2})^{-\frac{1}{2}} \exp\Big( -\frac{1}{2\sigma^{2}\tau_{ej}^{2}}\zeta_{j}^{2}\Big) \textbf{I}_{\{\zeta_{j} \neq 0\}}+ (1-\pi_{e})\delta_{0}(\zeta_{j}) \Bigg) \\
& \propto \pi_{e}^{r_{e}+\sum_{j=1}^{p}(\delta_{0}(\zeta_{j}))-1}(1-\pi_{e})^{w_{e}+\sum_{j=1}^{p}\textbf{I}_{\{\zeta_{j} \neq 0\}}-1}
\end{aligned}
\end{equation*}
the posterior distribution for $\pi_{e}$ is Beta($r_{e}+\sum_{j=1}^{p}(\delta_{0}(\zeta_{j}))$, $w_{e}+\sum_{j=1}^{p}\textbf{I}_{\{\zeta_{j} \neq 0\}}$).
Last, the full conditional distribution for $\sigma^{2}$
\begin{equation*}
\setlength{\jot}{10pt}
\begin{aligned}
\pi(\sigma^{2}| &\text{rest}) \\
\propto & \pi(\sigma^{2})\pi(y|\cdot)\prod_{j=1}^{p}\pi(\gamma_{j1}|\pi_{c},\tau_{cj}^{2},\sigma^{2})\pi(\gamma_{j*}|\pi_{v},\tau_{vj}^{2},\sigma^{2})\pi(\zeta_{j}|\pi_{e},\tau_{j}^{2},\sigma^{2}) \\
\propto & (\sigma^{2})^{-s-1}\exp(-\frac{h}{\sigma^{2}})(\sigma^{2})^{-\frac{n}{2}}\exp\Big(-\frac{1}{2\sigma^{2}}(Y-\mu)^\top (Y-\mu)\Big) \\
& \times \sum_{j=1}^{p} \Bigg( \pi_{c}(2\pi\sigma^{2}\tau_{cj}^{2})^{-\frac{1}{2}} \exp\Big( -\frac{1}{2\sigma^{2}\tau_{cj}^{2}}\gamma_{j1}^{2}\Big) \textbf{I}_{\{\gamma_{j1} \neq 0\}}+ (1-\pi_{c})\delta_{0}(\gamma_{j1}) \Bigg) \\
& \times \sum_{j=1}^{p} \Bigg( \pi_{v}(2\pi\sigma^{2}\tau_{vj}^{2})^{-\frac{L}{2}} \exp\Big( -\frac{1}{2\sigma^{2}\tau_{vj}^{2}}\gamma_{j*}^\top \gamma_{j*}\Big) \textbf{I}_{\{\gamma_{j*} \neq 0\}}+ (1-\pi_{v})\delta_{0}(\gamma_{j*}) \Bigg)\\
& \times \sum_{j=1}^{p} \Bigg( \pi_{e}(2\pi\sigma^{2}\tau_{ej}^{2})^{-\frac{1}{2}} \exp\Big( -\frac{1}{2\sigma^{2}\tau_{ej}^{2}}\zeta_{j}^{2}\Big) \textbf{I}_{\{\zeta_{j} \neq 0\}}+ (1-\pi_{e})\delta_{0}(\zeta_{j}) \Bigg) \\
\propto & (\sigma^{2})^{-(s+\frac{n+L\sum\textbf{I}_{\{\gamma_{j*}\neq0\}}+\sum\textbf{I}_{\{\zeta_{j} \neq 0\}}+\sum\textbf{I}_{\{\gamma_{j1} \neq 0\}}}{2})-1} \\
& \times \exp\Bigg(-\frac{1}{\sigma^{2}}\Big(h + \frac{(Y-\mu)^\top (Y-\mu) + \sum_{j=1}^{p}(\tau_{cj}^{2})^{-1}\gamma_{j1}^{2} + (\tau_{vj}^{2})^{-1}\gamma_{j*}^\top \gamma_{j*} + (\tau_{j}^{2})^{-1}\zeta_{j}^{2}}{2}\Big)\Bigg) 
\end{aligned}
\end{equation*}
the posterior distribution for $\sigma^{2}$ is Inverse-Gamma($\mu_{\sigma^{2}}$, $\Sigma_{\sigma^{2}}$) where
\begin{equation*}
\mu_{\sigma^{2}} = s+\frac{n+L\sum_{j=1}^{p}\textbf{I}_{\{\gamma_{j*}\neq0\}}+\sum_{j=1}^{p}\textbf{I}_{\{\zeta_{j} \neq 0\}}+\sum_{j=1}^{p}\textbf{I}_{\{\gamma_{j1} \neq 0\}}}{2}
\end{equation*}
and variance
\begin{equation*}
\Sigma_{\sigma^{2}} = h + \frac{(Y-\mu)^\top (Y-\mu) + \sum_{j=1}^{p}(\tau_{cj}^{2})^{-1}\gamma_{j1}^{2} + (\tau_{vj}^{2})^{-1}\gamma_{j*}^\top \gamma_{j*} + (\tau_{j}^{2})^{-1}\zeta_{j}^{2}}{2}
\end{equation*}

\subsection{Posterior inference for the BSSVC method}
\subsubsection{Priors}
\begin{equation*}
\begin{aligned}
Y|\eta, \gamma_{1},\dots,\gamma_{p},& \alpha_{1},\dots,\alpha_{q},\zeta_{0}, \zeta_{1},\dots,\zeta_{p},\sigma^{2} \\ 
& \propto (\sigma^{2})^{-\frac{n}{2}} \exp\Big\{-\frac{1}{2\sigma^{2}}(Y-\mu)^\top (Y-\mu)\Big\}
\end{aligned}
\end{equation*}

\begin{equation*}
\eta\thicksim \text{N}_{q_{n}}(0, \, \Sigma_{\eta0})
\end{equation*}
\begin{equation*}
\alpha\thicksim \text{N}_{q}(0, \, \Sigma_{\alpha0})
\end{equation*}
\begin{equation*}
\zeta_{0}\thicksim \text{N}(0, \, \sigma_{\zeta_{0}}^{2})
\end{equation*}
\begin{equation*}
\gamma_{j}|\pi_{c}, \tau_{vj}^{2}, \sigma^{2} \thicksim \pi_{v} \text{N}_{q_{n}}(0, \, \text{Diag}(\sigma^{2}\tau_{vj}^{2},\ldots, \sigma^{2}\tau_{vj}^{2})) + (1-\pi_{v})\delta_{0}(\gamma_{j}), \quad j=1,\dots, p
\end{equation*}
\begin{equation*}
\tau_{vj}^{2}|\lambda_{v} \thicksim \text{Gamma}(\frac{q_{n}+1}{2}, \, \frac{q_{n}\lambda_{v}^{2}}{2}), \quad j=1,\dots, p
\end{equation*}

\begin{equation*}
\zeta_{j}|\pi_{e}, \tau_{ej}^{2}, \sigma^{2} \thicksim \pi_{e} \text{N}(0, \, \sigma^{2}\tau_{ej}^{2})+ (1-\pi_{e})\delta_{0}(\zeta_{j}), \quad j=1,\dots, p
\end{equation*}
\begin{equation*}
\tau_{ej}^{2}|\lambda_{e} \thicksim \frac{\lambda_{e}^{2}}{2}\exp (-\frac{\lambda_{e}^{2}\tau_{ej}^{2}}{2}), \quad j=1,\dots, p
\end{equation*}

\begin{equation*}
\sigma^{2} \thicksim (\sigma^{2})^{-s-1}\exp(-\frac{h}{\sigma^{2}})
\end{equation*}

Consider the following conjugate gamma priors for $\lambda_{v}^{2}$ and $\lambda_{e}^{2}$
\begin{equation*}
\lambda_{v}^{2} \thicksim \text{Gamma}(a_{v}, \, b_{v}) \quad \text{and}\quad \lambda_{e}^{2} \thicksim \text{Gamma}(a_{e}, \, b_{e})
\end{equation*}

and conjugate beta priors for  $\pi_{v}$ and $\pi_{e}$
\begin{equation*}\label{equr:pi}
\pi_{v} \thicksim \text{Beta}(r_{v}, \, w_{v}) \quad \text{and}\quad \pi_{e} \thicksim \text{Beta}(r_{e}, \, w_{e})
\end{equation*}

\subsubsection{Posterior distribution}
$\pi(\eta|\text{rest}) \thicksim \text{N}_{q_{n}}(\mu_{\eta},\, \Sigma_{\eta})$ where 
\begin{equation*}
\setlength{\jot}{10pt}
\begin{aligned}
\mu_{\eta}&=\Big(\Sigma_{\eta0}^{-1}+ \frac{1}{\sigma^{2}}B_{0}^\top B_{0}\Big)^{-1} \Big(\frac{1}{\sigma^{2}}(Y-\mu_{(-\eta)})^\top B_{0}\Big)^\top \\
\Sigma_{\eta}&=\Big(\Sigma_{\eta0}^{-1}+ \frac{1}{\sigma^{2}}B_{0}^\top B_{0}\Big)^{-1}
\end{aligned}
\end{equation*}
$\pi(\alpha|\text{rest}) \thicksim \text{N}_{q}(\mu_{\alpha}, \Sigma_{\alpha})$ where 
\begin{equation*}
\setlength{\jot}{10pt}
\begin{aligned}
\mu_{\alpha}&=\Big(\Sigma_{\alpha0}^{-1}+ \frac{1}{\sigma^{2}}W^\top W\Big)^{-1} \Big(\frac{1}{\sigma^{2}}(Y-\mu_{(-\alpha)})^\top W\Big)^\top \\
\Sigma_{\alpha}&=\Big(\Sigma_{\alpha0}^{-1}+ \frac{1}{\sigma^{2}}W^\top W\Big)^{-1}
\end{aligned}
\end{equation*}
$\pi(\zeta_{0}|\text{rest}) \thicksim \text{N}(\mu_{\zeta_{0}}, \Sigma_{\zeta_{0}})$ where
\begin{equation*}
\setlength{\jot}{10pt}
\begin{aligned}
\mu_{\zeta_{0}}&=\Big(\sigma_{\zeta0}^{-1}+\frac{1}{\sigma^{2}}\sum_{i=1}^{n}E_{i}^{2}\Big)^{-1} \Big(\frac{1}{\sigma^{2}}\sum_{i=1}^{n}(y_{i}-\mu_{(-\zeta_{0})})E_{i}\Big) \\
\Sigma_{\zeta0}&=\Big(1/\sigma_{\zeta_{0}}^{2} + \frac{1}{\sigma^{2}}\sum_{i=1}^{n}E_{i}^{2}\Big)^{-1}
\end{aligned}
\end{equation*}
$\gamma_{j}|\text{rest} \thicksim l_{vj} \text{N}(\mu_{\gamma_{j}}, \, \sigma^{2}\Sigma_{\gamma_{j}}) + (1-l_{vj})\delta_{0}(\gamma_{j})$ where
\begin{equation*}
\setlength{\jot}{10pt}
\begin{aligned}
\mu_{\gamma_{j}} &= \Sigma_{\gamma_{j}}U_{j}^\top(Y - \mu_{(-\gamma_{j})})  \\
\Sigma_{\gamma_{j}} &= (U_{j}^\top U_{j} + \frac{1}{\tau_{vj}^{2}}\textbf{I}_{q_{n}})^{-1} \\
l_{vj} &= \frac{\pi_{v}}{\pi_{v} + (1-\pi_{v})(\tau_{vj}^{2})^{\frac{q_{n}}{2}}|\Sigma_{\gamma_{j}}|^{-\frac{1}{2}} \exp\Big(- \frac{1}{2\sigma^{2}} \lVert\Sigma_{\gamma_{j}}^{\frac{1}{2}}U_{j}^\top(Y-\mu_{(-\gamma_{j})})\rVert_{2}^{2} \Big)}
\end{aligned}
\end{equation*}
$\zeta_{j}|\text{rest} \thicksim l_{ej} \text{N}(\mu_{\zeta_{j}}, \, \sigma^{2}\Sigma_{\zeta_{j}}) + (1-l_{ej})\delta_{0}(\zeta_{j})$ where
\begin{equation*}
\setlength{\jot}{10pt}
\begin{aligned}
\mu_{\zeta_{j}} &=\Sigma_{\zeta_{j}}T_{j}^\top(Y - \mu_{(-\zeta_{j})})  \\
\Sigma_{\zeta_{j}} &= (T_{j}^\top T_{j} + \frac{1}{\tau_{ej}^{2}})^{-1} \\
l_{ej} &= \frac{\pi_{e}}{\pi_{e} + (1-\pi_{e})(\tau_{ej}^{2})^{\frac{1}{2}}(\Sigma_{\zeta_{j}})^{-\frac{1}{2}} \exp\Big(- \frac{\Sigma_{\zeta_{j}}}{2\sigma^{2}} \lVert(Y-\mu_{(-\zeta_{j})})^\top T_{j}\rVert_{2}^{2} \Big)} 
\end{aligned}
\end{equation*}
At the $g$th iteration, the values of $\phi_{vj}^{(g)}$ and $\phi_{ej}^{(g)}$ can be determined by whether the $\gamma_{j}^{(g)}$ and $\zeta_{j}^{(g)}$ are set to $0$ or not, respectively. 
\begin{equation*}
(\tau_{vj}^{2})^{-1}|\text{rest} \thicksim \begin{cases}
\scalebox{1}{Inverse-Gamma($\frac{q_{n}+1}{2}$,\, $\frac{q_{n}\lambda_{v}^{2}}{2}$)}& { \text{if} \; \gamma_{j} = 0} \\[5pt]
\scalebox{1}{Inverse-Gaussian($q_{n}\lambda_{v}^{2}$, $\sqrt{\frac{q_{n}\lambda_{v}^{2}\sigma^{2}}{\lVert\gamma_{j}\rVert_{2}^{2}}}$)}& { \text{if} \; \gamma_{j} \neq 0}
\end{cases}
\end{equation*}

\begin{equation*}
(\tau_{ej}^{2})^{-1}|\text{rest} \thicksim \begin{cases}
\scalebox{1}{Inverse-Gamma($1$,\, $\frac{\lambda_{e}^{2}}{2}$)}& { \text{if} \; \zeta_{j} = 0} \\[5pt]
\scalebox{1}{Inverse-Gaussian($\lambda_{e}^{2}$, $\sqrt{\frac{\lambda_{e}^{2}\sigma^{2}}{\zeta_{j}^{2}}}$)}& { \text{if} \; \zeta_{j} \neq 0}
\end{cases}
\end{equation*}
$\lambda_{v}$ and $\lambda_{e}$ all have inverse-gamma posterior distributions
\begin{equation*}
\setlength{\jot}{3pt}
\begin{aligned}
\lambda_{v}^{2} &\thicksim \text{Inverse-Gamma}(a_{v}+\frac{p(q_{n}+1)}{2}, \, b_{v}+\frac{q_{n}\sum_{j=1}^{p}\tau_{vj}^{2}}{2}) \\
\lambda_{e}^{2}&\thicksim \text{Inverse-Gamma}(a_{e}+p,\, b_{e}+\frac{\sum_{j=1}^{p}\tau_{ej}^{2}}{2})  \\
\end{aligned}
\end{equation*}
$\pi_{v}$ and $\pi_{e}$ have beta posterior distributions
\begin{equation*}
\setlength{\jot}{3pt}
\begin{aligned}
\pi_{v} &\thicksim \text{Beta}(r_{v}+\sum_{j=1}^{p}(\delta_{0}(\gamma_{j})), \, w_{v}+\sum_{j=1}^{p}\textbf{I}_{\{\gamma_{j} \neq 0\}}) \\
\pi_{e}&\thicksim \text{Beta}(r_{e}+\sum_{j=1}^{p}(\delta_{0}(\zeta_{j})), \, w_{e}+\sum_{j=1}^{p}\textbf{I}_{\{\zeta_{j} \neq 0\}})  \\
\end{aligned}
\end{equation*}
$\sigma^{2}\thicksim \text{Inverse-Gamma}(\mu_{\sigma^{2}}, \Sigma_{\sigma^{2}})$ where
\begin{equation*}
\setlength{\jot}{10pt}
\begin{aligned}
\mu_{\sigma^{2}} &= s+\frac{n+q_{n}\sum_{j=1}^{p}\textbf{I}_{\{\gamma_{j}\neq0\}}+\sum_{j=1}^{p}\textbf{I}_{\{\zeta_{j} \neq 0\}}}{2} \\
\Sigma_{\sigma^{2}} &= h + \frac{(Y-\mu)^\top (Y-\mu) + \sum_{j=1}^{p} (\tau_{vj}^{2})^{-1}\gamma_{j}^\top \gamma_{j} + (\tau_{cj}^{2})^{-1}\zeta_{j}^{2}}{2}
\end{aligned}
\end{equation*}

\subsection{Posterior inference for the BVC-SI method}
\subsubsection{Priors}
\begin{equation*}
\begin{aligned}
Y|\eta, \gamma_{11},\dots,\gamma_{p1},& \gamma_{1*},\dots,\gamma_{p*},\alpha_{1},\dots,\alpha_{q},\zeta_{0}, \zeta_{1},\dots,\zeta_{p},\sigma^{2} \\ 
& \propto (\sigma^{2})^{-\frac{n}{2}} \exp\Big\{-\frac{1}{2\sigma^{2}}(Y-\mu)^\top (Y-\mu)\Big\}
\end{aligned}
\end{equation*}
\begin{equation*}
\eta\thicksim \text{N}_{q_{n}}(0, \Sigma_{\eta0})
\end{equation*}
\begin{equation*}
\alpha\thicksim \text{N}_{q}(0, \Sigma_{\alpha0})
\end{equation*}
\begin{equation*}
\zeta_{0}\thicksim \text{N}(0, \, \sigma_{\zeta_{0}}^{2})
\end{equation*}
\begin{equation*}\label{equr:r0}
\gamma_{j1}|\tau_{cj}^{2}, \sigma^{2} \thicksim \text{N}(0, \, \sigma^{2}\tau_{cj}^{2}), \quad j=1,\dots, p
\end{equation*}
\begin{equation*}\label{equr:tau0}
\tau_{cj}^{2}|\lambda_{c} \thicksim \frac{\lambda_{c}^{2}}{2}\exp (-\frac{\lambda_{c}^{2}\tau_{cj}^{2}}{2}), \quad j=1,\dots, p
\end{equation*}
\begin{equation*}
\gamma_{j*}|\tau_{vj}^{2}, \sigma^{2} \thicksim \text{N}_{L}(0, \, \text{diag}(\sigma^{2}\tau_{vj}^{2},\ldots, \sigma^{2}\tau_{vj}^{2})), \quad j=1,\dots, p
\end{equation*}
\begin{equation*}
\tau_{vj}^{2}|\lambda_{v} \thicksim \text{Gamma}(\frac{L+1}{2}, \, \frac{L\lambda_{v}^{2}}{2}), \quad j=1,\dots, p
\end{equation*}
\begin{equation*}
\zeta_{j}|\tau_{ej}^{2}, \sigma^{2} \thicksim \text{N}(0, \, \sigma^{2}\tau_{ej}^{2}), \quad j=1,\dots, p
\end{equation*}
\begin{equation*}
\tau_{ej}^{2}|\lambda \thicksim \frac{\lambda_{e}^{2}}{2}\exp (-\frac{\lambda_{e}^{2}\tau_{ej}^{2}}{2}), \quad j=1,\dots, p
\end{equation*}

\begin{equation*}
\sigma^{2} \thicksim (\sigma^{2})^{-s-1}\exp(-\frac{h}{\sigma^{2}})
\end{equation*}

Consider the following conjugate gamma priors for $\lambda_{c}^2$, $\lambda_{v}^{2}$ and $\lambda_{e}^{2}$
\begin{equation*}
\lambda_{c}^{2} \thicksim \text{Gamma}(a_{c}, \, b_{c}), \quad \lambda_{v}^{2} \thicksim \text{Gamma}(a_{v}, \, b_{v}) \quad \text{and}\quad \lambda_{e}^{2} \thicksim \text{Gamma}(a_{e}, \, b_{e})
\end{equation*}

\subsubsection{Gibbs Sampler}
$\pi(\eta|\text{rest}) \thicksim \text{N}_{q_{n}}(\mu_{\eta},\, \Sigma_{\eta})$ where 
\begin{equation*}
\setlength{\jot}{10pt}
\begin{aligned}
\mu_{\eta}&=\Big(\Sigma_{\eta0}^{-1}+ \frac{1}{\sigma^{2}}B_{0}^\top B_{0}\Big)^{-1} \Big(\frac{1}{\sigma^{2}}(Y-\mu_{(-\eta)})^\top B_{0}\Big)^\top \\
\Sigma_{\eta}&=\Big(\Sigma_{\eta0}^{-1}+ \frac{1}{\sigma^{2}}B_{0}^\top B_{0}\Big)^{-1}
\end{aligned}
\end{equation*}
$\pi(\alpha|\text{rest}) \thicksim \text{N}_{q}(\mu_{\alpha}, \Sigma_{\alpha})$ where 
\begin{equation*}
\setlength{\jot}{10pt}
\begin{aligned}
\mu_{\alpha}&=\Big(\Sigma_{\alpha0}^{-1}+ \frac{1}{\sigma^{2}}W^\top W\Big)^{-1} \Big(\frac{1}{\sigma^{2}}(Y-\mu_{(-\alpha)})^\top W\Big)^\top \\
\Sigma_{\alpha}&=\Big(\Sigma_{\alpha0}^{-1}+ \frac{1}{\sigma^{2}}W^\top W\Big)^{-1}
\end{aligned}
\end{equation*}
$\pi(\zeta_{0}|\text{rest}) \thicksim \text{N}(\mu_{\zeta_{0}}, \Sigma_{\zeta_{0}})$ where
\begin{equation*}
\setlength{\jot}{10pt}
\begin{aligned}
\mu_{\zeta_{0}}&=\Big(1/\sigma_{\zeta0}^{2}+\frac{1}{\sigma^{2}}\sum_{i=1}^{n}E_{i}^{2}\Big)^{-1} \Big(\frac{1}{\sigma^{2}}\sum_{i=1}^{n}(y_{i}-\mu_{(-\zeta_{0})})E_{i}\Big) \\
\Sigma_{\zeta0}&=\Big(1/\sigma_{\zeta_{0}}^{2} + \frac{1}{\sigma^{2}}\sum_{i=1}^{n}E_{i}^{2}\Big)^{-1}
\end{aligned}
\end{equation*}
The full conditional distribution of $\gamma_{j}$ 
\begin{equation*}
\setlength{\jot}{10pt}
\begin{aligned}
\pi(\gamma_{j*}|& rest) \\
& \propto \pi(\gamma_{j*}|\tau_{vj}^{2}, \sigma^{2})\pi(y|\cdot) \\
& \propto \exp\Bigg(-\frac{1}{2}(\sigma^{2}\tau_{vj}^{2})^{-1}\gamma_{j*}^\top \gamma_{j*} -\frac{1}{2\sigma^{2}}(Y-U_{j}\gamma_{j*}-\mu_{(-\gamma_{j*})})^\top (Y-U_{j}\gamma_{j*}-\mu_{(-\gamma_{j*})})\Bigg) \\
& \propto \exp\Bigg(-\frac{1}{2\sigma^{2}}\Big((\tau_{vj}^{2})^{-1}\gamma_{j*}^\top \gamma_{j*} + \gamma_{j*}^\top U_{j}^\top U_{j}\gamma_{j*} - 2(Y-\mu_{(-\gamma_{j*})})^\top (U_{j}\gamma_{j*})\Big)\Bigg) \\
& \propto \exp\Bigg(-\frac{1}{2\sigma^{2}}\Big(\gamma_{j*}^\top \Big((\tau_{vj}^{2})^{-1} + U_{j}^\top U_{j}\Big)\gamma_{j*} -2(Y-\mu_{(-\gamma_{j*})})^\top U_{j}\gamma_{j*}\Big)\Bigg)
\end{aligned}
\end{equation*}
where $U_{j} = (U_{1j}, \dots, U_{nj})^\top $. Hence, the full conditional distribution of $\gamma_{j*}$ is multivariate normal with mean
\begin{equation*}
\mu_{\gamma_{j*}}=\Big((\tau_{vj}^{2})^{-1}\textbf{I}_{L} + U_{j}^\top U_{j}\Big)^{-1} \Big((Y-\mu_{(-\gamma_{j*})})^\top U_{j}\Big)^\top 
\end{equation*}
and variance
\begin{equation*}
\Sigma_{\gamma_{j*}}=\sigma^{2}\Big((\tau_{vj}^{2})^{-1}\textbf{I}_{L} + U_{j}^\top U_{j}\Big)^{-1}
\end{equation*}
Similarly, the full conditional distribution of $\gamma_{j1}$ is normal distribution with mean
\begin{equation*}\label{equr:meanr}
\mu_{\gamma_{j1}}=\Big((\tau_{cj}^{2})^{-1}+ X_{j}^\top X_{j}\Big)^{-1} \Big((y_{i}-\mu_{(-\gamma_{j1})})^\top X_{j}\Big)
\end{equation*}
and variance
\begin{equation*}\label{equr:varr}
\Sigma_{\gamma_{j1}}=\sigma^{2}\Big((\tau_{cj}^{2})^{-1}+ X_{j}^\top X_{j}\Big)^{-1} 
\end{equation*}
Let $D_{\tau_{e}}=\text{diag}(\tau_{e1}^{2},\ldots,\tau_{ep}^{2})$. The full conditional distribution of $\zeta = (\zeta_{1},\dots,\zeta_{p})^\top $ 
\begin{equation*}\label{equr:posr}
\setlength{\jot}{10pt}
\begin{aligned}
\pi(\zeta|& rest) \\
& \propto \pi(\zeta|\tau_{e1}^{2},\ldots,\tau_{ep}^{2}, \sigma^{2} )\pi(y|\cdot) \\
& \propto \exp\Bigg(-\frac{1}{2}(\sigma^{2}D_{\tau_{e}})^{-1}\zeta^\top \zeta -\frac{1}{2\sigma^{2}}(Y-T\zeta-\mu_{(-\zeta)})^\top (Y-T\zeta-\mu_{(-\zeta)})\Bigg) \\
& \propto \exp\Bigg(-\frac{1}{2\sigma^{2}}\Big(D_{\tau_{e}}^{-1}\zeta^\top \zeta + \zeta^\top T^\top T\zeta - 2(Y-\mu_{(-\zeta)})^\top T\zeta\Big)\Bigg) \\
& \propto \exp\Bigg(-\frac{1}{2\sigma^{2}}\Big(\zeta^\top (D_{\tau_{e}}^{-1} + T^\top T)\zeta -2(Y-\mu_{(-\zeta)})^\top T\zeta\Big)\Bigg)
\end{aligned}
\end{equation*}
where $T = (T_{1}, \dots, T_{n})^\top $. The full conditional is $\text{N}_{p}(\mu_{\zeta}, \sigma^{2}\Sigma_{\zeta})$ with
\begin{equation*}\label{equr:meana}
\mu_{\zeta}=\Big(D_{\tau_{e}}^{-1} + T^\top T\Big)^{-1} \Big((Y-\mu_{(-\zeta)})^\top T\Big)^\top 
\end{equation*}
and variance
\begin{equation*}\label{equr:vara}
\Sigma_{\zeta}=\Big(D_{\tau_{e}}^{-1} + T^\top T\Big)^{-1}
\end{equation*}
Now, we derive the full conditional distribution for $\tau_{ej}^{2}$ and $\lambda_{e}^{2}$.
\begin{equation*}\label{equr:postau}
\setlength{\jot}{10pt}
\begin{aligned}
\pi(\tau_{vj}^{2}|& rest) \\
& \propto \pi(\tau_{vj}^{2}|\lambda_{v}) \pi(\gamma_{j*}|\tau_{vj}^{2}, \sigma^{2}) \\
& \propto (\tau_{vj}^{2})^{\frac{L+1}{2}-1}\exp\Bigg(-\tau_{vj}^{2}\frac{L\lambda_{v}^{2}}{2}\Bigg) (\tau_{vj}^{2})^{-\frac{L}{2}}\exp\Bigg(-\frac{1}{2}(\sigma^{2}\tau_{vj}^{2})^{-1}\gamma_{j*}^\top \gamma_{j*}\Bigg) \\
& \propto (\tau_{vj}^{2})^{-\frac{1}{2}}\exp\Bigg(-\tau_{vj}^{2}\frac{L\lambda_{v}^{2}}{2} -\frac{||\gamma_{j*}||_{2}^{2}}{2\sigma^{2}\tau_{vj}^{2}}\Bigg) \\
\end{aligned}
\end{equation*}
the posterior distribution for $(\tau_{vj}^{2})^{-1}$ is Inverse-Gaussian($L\lambda_{v}^{2}$, $\sqrt{\frac{L\lambda_{v}^{2}\sigma^{2}}{||\gamma_{j*}||_{2}^{2}}}$). Similarly, the posterior distribution for $(\tau_{cj}^{2})^{-1}$ is Inverse-Gaussian($\lambda_{c}^{2}$, $\sqrt{\frac{\lambda_{c}^{2}\sigma^{2}}{\gamma_{j1}^{2}}}$), and the posterior distribution for $(\tau_{ej}^{2})^{-1}$ is Inverse-Gaussian($\lambda_{e}^{2}$, $\sqrt{\frac{\lambda_{e}^{2}\sigma^{2}}{\zeta_{j}^{2}}}$).
\begin{equation*}
\setlength{\jot}{10pt}
\begin{aligned}
\pi(\lambda_{v}^{2}|& rest) \\
& \propto \pi(\lambda_{v}^{2})\prod_{j=1}^{p}\pi(\tau_{vj}^{2}|\lambda_{v}^{2}) \\
& \propto (\lambda_{v}^{2})^{a_{v}-1}\exp(-b_{v}\lambda_{v}^{2} )\prod_{j=1}^{p}\Bigg(\frac{L\lambda_{v}^{2}}{2}\Bigg)^{\frac{L+1}{2}}\exp\Bigg(-\frac{L\lambda_{v}^{2}}{2}\tau_{vj}^{2}\Bigg)\\
& \propto (\lambda_{v}^{2})^{a_{v}+\frac{p(L+1)}{2}-1} \exp\Bigg(-(b_{v}+\frac{L\sum_{j=1}^{p}\tau_{vj}^{2}}{2})\lambda_{v}^{2}\Bigg)
\end{aligned}
\end{equation*}
the posterior distribution for $\lambda_{c}^{2}$ is Inverse-Gamma($a_{c}+p$, $b_{c}+\frac{\sum_{j=1}^{p}\tau_{cj}^{2}}{2}$). Similarly, the full conditional distribution for $\lambda_{e}^{2}$ is Inverse-Gamma($a_{e}+p$, $b_{e}+\frac{\sum_{j=1}^{p}\tau_{ej}^{2}}{2}$).

Last, the full conditional distribution for $\sigma^{2}$
\begin{equation*}
\setlength{\jot}{10pt}
\begin{aligned}
\pi(\sigma^{2}|& rest) \\
& \propto \pi(\sigma^{2})\pi(y|\cdot)\prod_{j=1}^{p}\pi(\gamma_{j1}|\pi_{c},\tau_{cj}^{2},\sigma^{2})\pi(\gamma_{j*}|\pi_{v},\tau_{vj}^{2},\sigma^{2})\pi(\zeta_{j}|\pi_{e},\tau_{ej}^{2},\sigma^{2}) \\
& \propto (\sigma^{2})^{-s-1}\exp(-\frac{h}{\sigma^{2}})(\sigma^{2})^{-\frac{n}{2}}\exp\Big(-\frac{1}{2\sigma^{2}}(Y-\mu)^\top (Y-\mu)\Big) \\
&\quad\quad \times (\sigma^{2})^{-\frac{p}{2}}\exp\Big(-\frac{1}{2\sigma^{2}}\sum_{j=1}^{p}(\tau_{cj}^{2})^{-1}\gamma_{j1}^{2}\Big)\\
&\quad\quad \times (\sigma^{2})^{-\frac{pL}{2}}\exp\Big(-\frac{1}{2\sigma^{2}}\sum_{j=1}^{p}(\tau_{vj}^{2})^{-1}\gamma_{j*}^\top \gamma_{j*}\Big) \\
&\quad\quad \times
(\sigma^{2})^{-\frac{p}{2}}\exp\Big(-\frac{1}{2\sigma^{2}}\sum_{j=1}^{p}(\tau_{ej}^{2})^{-1}\zeta_{j}^{2}\Big) \\
& \propto (\sigma^{2})^{-(s+\frac{n+2p+pL}{2})-1} \\ 
& \quad\quad \times \exp\Bigg(-\frac{1}{\sigma^{2}}\Big(h+\frac{(Y-\mu)^\top (Y-\mu) + \sum_{j=1}^{p}(\tau_{cj}^{2})^{-1}\gamma_{j1}^{2} + (\tau_{vj}^{2})^{-1}\gamma_{j*}^\top \gamma_{j*} + (\tau_{ej}^{2})^{-1}\zeta_{j}^{2}}{2}\Big)\Bigg) 
\end{aligned}
\end{equation*}
the posterior distribution for $\sigma^{2}$ is Inverse-Gamma($\mu_{\sigma^{2}}$, $\Sigma_{\sigma^{2}}$) where
\begin{equation*}
\mu_{\sigma^{2}} = s+\frac{n+2p+pL}{2}
\end{equation*}
and variance
\begin{equation*}
\Sigma_{\sigma^{2}} = h + \frac{(Y-\mu)^\top (Y-\mu) + \sum_{j=1}^{p}(\tau_{cj}^{2})^{-1}\gamma_{j1}^{2} + (\tau_{vj}^{2})^{-1}\gamma_{j*}^\top \gamma_{j*} + (\tau_{ej}^{2})^{-1}\zeta_{j}^{2} }{2}
\end{equation*}

\subsection{Posterior inference for the BVC method}
\label{makereferenceA.10}
\subsubsection{Priors}
\begin{equation*}\label{equr:LKH}
Y|\eta, \gamma_{1},\dots,\gamma_{p},\alpha_{1},\dots,\alpha_{q},\zeta_{0},\zeta_{1},\dots,\zeta_{p},\sigma^{2} \propto (\sigma^{2})^{-\frac{n}{2}} \exp\Big\{-\frac{1}{2\sigma^{2}}(Y-\mu)^\top (Y-\mu)\Big\}
\end{equation*}

\begin{equation*}\label{equr:eta}
\eta\thicksim \text{N}_{q_{n}}(0, \Sigma_{\eta0})
\end{equation*}
\begin{equation*}\label{equr:alpha}
\alpha\thicksim \text{N}_{q}(0, \Sigma_{\alpha0})
\end{equation*}
\begin{equation*}
\zeta_{0}\thicksim \text{N}(0, \, \sigma_{\zeta_{0}}^{2})
\end{equation*}

\begin{equation*}\label{equr:pr}
\gamma_{j}|\tau_{vj}^{2}, \sigma^{2} \thicksim \text{N}_{q_{n}}(0, \, \text{diag}(\sigma^{2}\tau_{vj}^{2},\ldots, \sigma^{2}\tau_{vj}^{2})), \quad j=1,\dots, p
\end{equation*}
\begin{equation*}\label{equr:taustar}
\tau_{vj}^{2}|\lambda_{v} \thicksim \text{Gamma}(\frac{q_{n}+1}{2}, \, \frac{q_{n}\lambda_{v}^{2}}{2}), \quad j=1,\dots, p
\end{equation*}

\begin{equation*}\label{equr:zeta}
\zeta_{j}|\tau_{ej}^{2}, \sigma^{2} \thicksim N(0, \, \sigma^{2}\tau_{ej}^{2}), \quad j=1,\dots, p
\end{equation*}
\begin{equation*}\label{equr:tau}
\tau_{ej}^{2}|\lambda_{e} \thicksim \frac{\lambda_{e}^{2}}{2}\exp (-\frac{\lambda_{e}^{2}\tau_{ej}^{2}}{2}), \quad j=1,\dots, p
\end{equation*}

\begin{equation*}\label{equr:sig}
\sigma^{2} \thicksim (\sigma^{2})^{-s-1}\exp(-\frac{h}{\sigma^{2}})
\end{equation*}

Consider the following conjugate gamma priors for $\lambda_{v}^{2}$ and $\lambda_{e}^{2}$
\begin{equation*}\label{equr:ptau}
\lambda_{v}^{2} \thicksim \text{Gamma}(a_{v}, \, b_{v}) \quad \text{and}\quad \lambda_{e}^{2} \thicksim \text{Gamma}(a_{e}, \, b_{e})
\end{equation*}

\subsubsection{Gibbs Sampler}
$\pi(\eta|\text{rest}) \thicksim \text{N}_{q_{n}}(\mu_{\eta},\, \Sigma_{\eta})$ where 
\begin{equation*}
\setlength{\jot}{10pt}
\begin{aligned}
\mu_{\eta}&=\Big(\Sigma_{\eta0}^{-1}+ \frac{1}{\sigma^{2}}B_{0}^\top B_{0}\Big)^{-1} \Big(\frac{1}{\sigma^{2}}(Y-\mu_{(-\eta)})^\top B_{0}\Big)^\top \\
\Sigma_{\eta}&=\Big(\Sigma_{\eta0}^{-1}+ \frac{1}{\sigma^{2}}B_{0}^\top B_{0}\Big)^{-1}
\end{aligned}
\end{equation*}
$\pi(\alpha|\text{rest}) \thicksim \text{N}_{q}(\mu_{\alpha}, \Sigma_{\alpha})$ where 
\begin{equation*}
\setlength{\jot}{10pt}
\begin{aligned}
\mu_{\alpha}&=\Big(\Sigma_{\alpha0}^{-1}+ \frac{1}{\sigma^{2}}W^\top W\Big)^{-1} \Big(\frac{1}{\sigma^{2}}(Y-\mu_{(-\alpha)})^\top W\Big)^\top \\
\Sigma_{\alpha}&=\Big(\Sigma_{\alpha0}^{-1}+ \frac{1}{\sigma^{2}}W^\top W\Big)^{-1}
\end{aligned}
\end{equation*}
$\pi(\zeta_{0}|\text{rest}) \thicksim \text{N}(\mu_{\zeta_{0}}, \Sigma_{\zeta_{0}})$ where
\begin{equation*}
\setlength{\jot}{10pt}
\begin{aligned}
\mu_{\zeta_{0}}&=\Big(1/\sigma_{\zeta0}^{2}+\frac{1}{\sigma^{2}}\sum_{i=1}^{n}E_{i}^{2}\Big)^{-1} \Big(\frac{1}{\sigma^{2}}\sum_{i=1}^{n}(y_{i}-\mu_{(-\zeta_{0})})E_{i}\Big) \\
\Sigma_{\zeta_{0}}&=\Big(1/\sigma_{\zeta_{0}}^{2} + \frac{1}{\sigma^{2}}\sum_{i=1}^{n}E_{i}^{2}\Big)^{-1}
\end{aligned}
\end{equation*}
$\gamma_{j}|\text{rest} \thicksim \text{N}_{q_{n}}(\mu_{\gamma_{j}}, \, \sigma^{2}\Sigma_{\gamma_{j}})$ where
\begin{equation*}
\setlength{\jot}{10pt}
\begin{aligned}
\mu_{\gamma_{j}} &= \Sigma_{\gamma_{j}}U_{j}^\top(Y - \mu_{(-\gamma_{j})})  \\
\Sigma_{\gamma_{j}} &= (U_{j}^\top U_{j} + \frac{1}{\tau_{vj}^{2}}\textbf{I}_{q_{n}})^{-1} 
\end{aligned}
\end{equation*}
$\zeta_{j}|\text{rest} \thicksim \text{N}(\mu_{\zeta_{j}}, \, \sigma^{2}\Sigma_{\zeta_{j}})$ where
\begin{equation*}
\setlength{\jot}{10pt}
\begin{aligned}
\mu_{\zeta_{j}} &=\Sigma_{\zeta_{j}}T_{j}^\top(Y - \mu_{(-\zeta_{j})})  \\
\Sigma_{\zeta_{j}} &= (T_{j}^\top T_{j} + \frac{1}{\tau_{ej}^{2}})^{-1}
\end{aligned}
\end{equation*}
The posterior distribution for $(\tau_{vj}^{2})^{-1}$ is Inverse-Gaussian($q_{n}\lambda_{v}^{2}$, $\sqrt{\frac{q_{n}\lambda_{v}^{2}\sigma^{2}}{||\gamma_{j}||_{2}^{2}}}$). Similarly, the posterior distribution for $(\tau_{ej}^{2})^{-1}$ is Inverse-Gaussian($\lambda_{e}^{2}$, $\sqrt{\frac{\lambda_{e}^{2}\sigma^{2}}{\zeta_{j}^{2}}}$).
$\lambda_{v}$ and $\lambda_{e}$ all have inverse-gamma posterior distributions
\begin{equation*}
\setlength{\jot}{3pt}
\begin{aligned}
\lambda_{v}^{2} &\thicksim \text{Inverse-Gamma}(a_{v}+\frac{p(q_{n}+1)}{2}, \, b_{v}+\frac{q_{n}\sum_{j=1}^{p}\tau_{vj}^{2}}{2}) \\
\lambda_{e}^{2}&\thicksim \text{Inverse-Gamma}(a_{e}+p,\, b_{e}+\frac{\sum_{j=1}^{p}\tau_{ej}^{2}}{2})  \\
\end{aligned}
\end{equation*}
$\sigma^{2}\thicksim \text{Inverse-Gamma}(\mu_{\sigma^{2}}, \Sigma_{\sigma^{2}})$ where
\begin{equation*}
\setlength{\jot}{10pt}
\begin{aligned}
\mu_{\sigma^{2}} &= s + \frac{n+p+pq_{n}}{2} \\
\Sigma_{\sigma^{2}} &= h + \frac{1}{2}\Big((Y-\mu)^\top(Y-\mu) + \sum_{j=1}^{p} (\tau_{vj}^{2})^{-1}\gamma_{j}^\top\gamma_{j} + (\tau_{ej}^{2})^{-1}\zeta_{j}^{2} \Big)
\end{aligned}
\end{equation*}

\clearpage

%

\end{document}